\documentclass{IEEEoj}
\usepackage{graphicx,subfigure, siunitx}
\graphicspath{{./figures/},{./imgs/}}
\usepackage{booktabs,multirow,longtable}
\usepackage{algorithm,algorithmic}
\usepackage{balance}
\usepackage[cmex10]{amsmath}
\usepackage[utf8x]{inputenc}
\usepackage[T1]{fontenc}
\usepackage[american]{babel}
\usepackage{stfloats}
\usepackage{amsthm, amssymb, bm, bbm, graphicx, ucs, cite, relsize, booktabs,color, mathalfa, upgreek, tablefootnote, multirow, colortbl,tabularx}
\usepackage{anyfontsize}
\definecolor{Gray}{gray}{0.9}

\theoremstyle{remark}

\theoremstyle{definition}

\theoremstyle{remark}

\newcommand*{\herm}{^{\mathsf{H}}}
\newcommand*{\transp}{^{\mathsf{T}}}

\DeclareMathOperator{\diag}{diag}

\DeclareMathOperator*{\argmin}{\arg\min}
\DeclareMathOperator*{\argmax}{\arg\max}

\newcommand{\e}{\mathrm{e}}
\renewcommand{\i}{\mathrm{i}}

\IEEEoverridecommandlockouts
\allowdisplaybreaks

\newcommand{\test}{\mathrel{\underset{{\mathcal H}_{0}}{\overset{{\mathcal{H}}_{1}}{\gtrless}}}} 

\def\BibTeX{{\rm B\kern-.05em{\sc i\kern-.025em b}\kern-.08em
    T\kern-.1667em\lower.7ex\hbox{E}\kern-.125emX}}
\AtBeginDocument{\definecolor{ojcolor}{cmyk}{0.93,0.59,0.15,0.02}}
\def\OJlogo{\vspace{-4pt}\hskip-4pt\includegraphics[height=18pt]{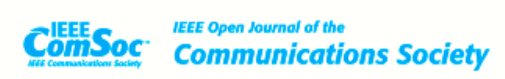}}
\allowdisplaybreaks[4]
\begin{document}

\receiveddate{February, 2025}
\reviseddate{March, 2025}
\accepteddate{May, 2025}
\publisheddate{TBD, 2025}
\currentdate{May, 2025}

\title{Integrated Communication and RIS-aided Track-Before-Detect Radar Sensing}


\author{Georgios Mylonopoulos\IEEEauthorrefmark{1},
Luca Venturino\IEEEauthorrefmark{1},~\IEEEmembership{Senior Member,~IEEE,}
Emanuele Grossi\IEEEauthorrefmark{1},~\IEEEmembership{Senior Member,~IEEE,}
Stefano Buzzi\IEEEauthorrefmark{1,2},~\IEEEmembership{Senior Member,~IEEE,} and
Ciro D'Elia\IEEEauthorrefmark{1}}
\affil{Department of Electrical and Information Engineering (DIEI), University of Cassino and Southern Lazio, 03043 Cassino, Italy and Consorzio Nazionale Interuniversitario per le Telecomunicazioni (CNIT), 43124 Parma, Italy.}
\affil{ Dipartimento di Elettronica
Informazione e Bioingegneria, Politecnico di Milano, Milan, Italy.}
\corresp{CORRESPONDING AUTHOR: Georgios Mylonopoulos (e-mail: georgios.mylonopoulos@unicas.it).}
\authornote{The work of G. Mylonopoulos and S. Buzzi was supported by the EU Horizon 2020 MSCA-ITN-METAWIRELESS, Grant Agreement 956256. The work of L. Venturino was supported by the project ``FLexible And distributed cognitive Radar systEms'' (FLARE), CUP E63C22002040007, funded by the European Union under the Italian National Recovery and Resilience Plan of NextGenerationEU, partnership on ``Telecommunications of the Future'' (Project PE00000001, program ``RESTART''). The work of E. Grossi was supported by the project ``CommunIcations and Radar Co-Existence (CIRCE),'' CUP H53D23000420006, funded by the European Union under the Italian National Recovery and Resilience Plan of NextGenerationEU. This work is accepted for publication on OJCOMS with Digital Object Identifier: 10.1109/OJCOMS.2025.3572081}

\begin{abstract}
This paper investigates an integrated sensing and communication system where the base station serves multiple downlink users, while employing a passive reconfigurable intelligent surface to detect small, noncooperative airborne targets. We propose a method to design the two-way beampattern of the RIS-assisted monostatic radar, which allows controlling the sidelobe levels in the presence of eavesdroppers, jammers, and other scattering objects and avoiding any radar interference to the users. To obtain more favorable system tradeoffs, we exploit the correlation of the target echoes over consecutive scans by resorting to a multi-frame radar detector, which includes a detector, a plot-extractor, and a track-before-detect processor. A numerical analysis is provided to verify the effectiveness of the proposed solutions and to assess the achievable tradeoffs. Our results show that, by increasing the number of scans processed by the radar detector (and therefore its implementation complexity), we can reduce the amount of power dedicated to the radar function while maintaining the same sensing performance (measured in terms of probability of target detection and root mean square error in the estimation of target position); this excess power can be reused to increase the user sum-rate.
\end{abstract}

\begin{IEEEkeywords}
Integrated sensing and communication (ISAC), reconfigurable intelligent surface (RIS), track-before-detect (TBD), multi-frame detection, orthogonal frequency division multiplexing (OFDM).
\end{IEEEkeywords}

\maketitle

\section{Introduction}
Integrated Sensing and Communication (ISAC) has emerged as a key paradigm for next-generation wireless networks, offering the potential for efficient resource utilization by sharing time, frequency, and spatial domains between sensing and communication functionalities~\cite{9606831}. This integration promises to revolutionize various applications, leveraging the dense deployment of the 5G infrastructure to improve sensing capabilities~\cite{9737357}. A central motivation for ISAC lies in the seamless integration of hardware and spectrum, minimizing modifications to existing infrastructure. However, realizing the full potential of ISAC requires addressing fundamental challenges, including the inherent performance tradeoffs between sensing and communication, as well as the diverse signaling requirements and evaluation metrics for each function~\cite{9705498, 10012421}. Effective resource management is crucial, especially in multi-user scenarios, where the interplay between sensing performance and communication Quality of Service (QoS) creates a complex optimization framework~\cite{9945983,10418473}. While existing studies have explored ISAC resource allocation, they often adopt a communication-centric perspective, focusing on optimizing specific sensing metrics under communication constraints, leading to complex non-convex problems~\cite{9965407,10623531}. These approaches often neglect the potential for adaptive resource management in spatially aware ISAC systems~\cite{10694524,9557830} and the rich set of sensing-centric signal processing techniques that can significantly enhance sensing performance~\cite{9540344}.

Reconfigurable Intelligent Surfaces (RISs) have become a focal point in ISAC research due to their ability to dynamically shape the propagation environment and extend the coverage of the system~\cite{9765815}. Integrating RISs into ISAC systems offers a new dimension in system design, as RIS placement influences available sensing points, thus improving coverage and resolution~\cite{10736517, 9737357}. However, balancing the performance tradeoffs between sensing and communication in RIS-aided ISAC systems is a critical concern~\cite{10077119}. Existing approaches often treat the RIS as an extension of the base station (BS), focusing on the joint beampattern design for both user coverage and sensing~\cite{9852716, 9844707, 10052711, 10792983}; these joint designs can become increasingly complex, leading to sequential design strategies in which BS and RIS configurations are optimized separately~\cite{9844707,10052711,10792983}. 
A key challenge in RIS-aided systems is obtaining accurate channel state information (CSI) for all RIS-to-user channels, which requires more elaborated training procedures~\cite{10466748,10156858,10025392}. Another challenge is the multiplicative path loss in the indirect link generated by the RIS, which may be mitigated by deploying the RIS closer to the BS and/or by resorting to active elements that amplify the reflected signal~\cite{10915665}.

In radar applications, Track-Before-Detect (TBD) is a powerful and consolidated technique for detecting weak moving targets~\cite{Boers_2004, Pulford_2010, Blanding_2007}; see also~\cite{Davey_2008} for a survey. TBD processes multiple scans (or frames) effectively accumulating the signal energy along the (unknown) target trajectory and improving detection sensitivity. While advanced TBD techniques can enhance performance in dense target environments, they also increase the computational complexity. To alleviate this problem,~\cite{Grossi-2013a, Grossi-2013b, Aprile-2016} adopts a two-stage approach, where the first stage limits the number of observations by retaining only the most significant ones and the second stage performs a dynamic trajectory formation to validate each detection. Recent studies have explored similar multi-frame processing techniques in dual-function sensing and communication systems, demonstrating improved detection of weak moving targets~\cite{ji2024modified}. However, the potential to exploit the temporal correlation of target echoes within the broader ISAC framework remains largely unexplored.

\subsection{Motivations and Contributions}
Driven by the growing demand to integrate sensing and communication capabilities, this paper explores an OFDM-based ISAC system where a BS serves multiple ground users, while a nearby RIS is exploited for detecting non-cooperative airborne targets. The main contributions of our work are summarized below.
\begin{itemize}
    \item We integrate the multi-frame TBD processor from~\cite{Grossi-2013a, Grossi-2013b, Aprile-2016} into the ISAC transceiver design in order to exploit the temporal correlation of the echoes generated by a moving target.
    \item We define the two-way beampattern of the considered RIS-assisted radar,  which specifically accounts for  the underlying OFDM-based signal structure and the effects of both the transmit and receive beampatterns. 
    \item We propose a method  to design the two-way beampattern, that maximizes the expected power for any target echo within the inspected subvolume, while controlling the transmit and/or receive sidelobe levels in the presence of eavesdroppers and/or interference sources and avoiding any impact on the communication function.
    \item Finally, we provide numerical results based on realistic 5G system specifications. Remarkably, our results demonstrate that increasing the number of scans processed by the radar detector (i.e., increasing its implementation complexity) allows for reduced power dedicated to the radar function, while maintaining the same sensing performance (measured in terms of probability of target detection and root mean square error in the estimation of the target position); this excess power can then be redirected to the communication function, so as to increase the user sum-rate. 
\end{itemize}

Previous related studies have already investigated several strategies for designing the downlink spatial beamformers of the BS~\cite{9591331,9729741,9858656,10158711,9668964,10086626}. In particular, relevant performance metrics for the communication function are the signal-to-interference-plus-noise-ratio (SINR) and the rate of the users, while the signal-to-noise-plus-clutter ratio, the Cramér Rao lower bound on the estimation of the unknown target parameters, and the transmit beampattern are common metrics for the sensing function. Common design criteria aim to maximize a communication metric under a constraint on a sensing metric~\cite{9591331,9729741,9858656,10158711} or vice-versa~\cite{9668964,10086626}.  This paper significantly differs from previous studies in the implementation of the sensing function, as outlined is Table~\ref{tab:lit_rev}. To the best of our knowledge, this is the first work to incorporate a multi-frame radar processing into an ISAC transceiver, recognizing it as a means to obtain more favorable sensing and communication tradeoffs. Also, for given downlink communication beamformers (that can be obtained for example from the aforementioned existing methods), we show that the RIS response and the sensing beamformer used by the BS to illuminate the RIS can be jointly designed to scan the volume of interest while controlling the sidelobes of the transmit and receive beampatterns and avoiding any interference on the communication function.

\begin{table*}
	\caption{Implementation of the sensing function in related ISAC studies}\label{tab:lit_rev}
    \begin{center}
    \begin{tabular}{c c c c  c c c  c}
    \toprule
                             &      \multicolumn{3}{c}{Receiver}         & \multicolumn{4}{c}{Beamforming}            \\   
         \cmidrule(lr){2-4} \cmidrule(lr){5-8} 
                                       & Multi-frame  & Target       & Target       & Transmit Beampattern & Receive Beampattern  &   Sidelobe   & \multirow{2}{*}{RIS-aided} \\
                                       & Processing   & Detection    & Localization &  Design              &   Design             &   Control    &                            \\
        \midrule
        \cite{9557830}                 & $\times$     & $\times$     & $\checkmark$ & $\checkmark$         & $\checkmark$         & $\times$     & $\times$                   \\
        \rowcolor{Gray}\cite{9852716}  & $\times$     & $\times$     & $\times$     & $\checkmark$         & $\times$             & $\checkmark$ & $\checkmark$               \\
        \cite{9844707}                 & $\times$     & $\times$     & $\times$     & $\checkmark$         & $\times$             & $\times$     & $\checkmark$               \\
        \rowcolor{Gray}\cite{10052711} & $\times$     & $\times$     & $\times$     & $\checkmark$         & $\times$             & $\times$     & $\checkmark$               \\
        \cite{10792983}                & $\times$     & $\times$     & $\times$     & $\checkmark$         & $\times$             & $\times$     & $\checkmark$                    \\
        \rowcolor{Gray}\cite{10915665} & $\times$     & $\times$     & $\times$     & $\checkmark$         & $\times$             & $\times$     & $\checkmark$                   \\
        \cite{9591331}                 & $\times$     & $\times$     & $\checkmark$ & $\checkmark$         & $\times$             & $\times$     & $\checkmark$               \\
        \rowcolor{Gray}\cite{9729741}  & $\times$     & $\times$     & $\times$     & $\checkmark$         & $\checkmark$         & $\times$     & $\checkmark$               \\
        \cite{9858656}                 & $\times$     & $\times$     & $\times$     & $\checkmark$         & $\times$             & $\times$     & $\times$                   \\
        \rowcolor{Gray}\cite{10158711} & $\times$     & $\checkmark$ & $\times$     & $\checkmark$         & $\checkmark$         & $\checkmark$ & $\times$                   \\
        \cite{9668964}                 & $\times$     & $\times$     & $\times$     & $\checkmark$         & $\times$             & $\times$     & $\times$                   \\
        \rowcolor{Gray}\cite{10086626} & $\times$     & $\times$     & $\times$     & $\checkmark$         & $\times$             & $\checkmark$ & $\times$                   \\
        This paper                     & $\checkmark$ & $\checkmark$ & $\checkmark$ & $\checkmark$         & $\checkmark$         & $\checkmark$ & $\checkmark$               \\
        \bottomrule
    \end{tabular}
    \end{center}
\end{table*}

\subsection{Organization} The remainder of this work is organized as follows. Sec.~\ref{sec:system} presents the system description and the underlying design assumptions. Sec.~\ref{sec:system_design} illustrates the proposed beampattern design and the incorporation of the TBD processor into the RIS-assisted radar transceiver. Sec.~\ref{sec:Numerical analysis} contains some numerical examples exploring the achievable tradeoffs between the communication and sensing performance. Finally, concluding remarks are given in Sec.~\ref{sec:conclusions}.

\subsection{Notation} Column vectors and matrices are denoted by lowercase and uppercase boldface letters, respectively. The symbols $(\cdot)\transp$, $(\cdot)^*$, $(\cdot)\herm$ and $(\cdot)^{+}$ denote the transpose, conjugate, conjugate-transpose and pseudo-inverse operations, respectively. $\bm{I}_M$ is the $M \times M$ identity matrix, while $\bm{0}_{M}$ is an $M \times 1$ vector of all zero elements. $\| \bm{\alpha}\|$ is the Euclidean norm of the vector $\bm{\alpha}$ and $\text{diag}\{\bm{\alpha}\}$ is the $N \times N$ diagonal matrix with the entries of vector $ \bm{\alpha} $ on the main diagonal. $[\bm{A}]_{i,j}$ marks the element in the $i$-th row and $j$-th column for matrix $\bm{A}$. Finally, the symbols $\i$, $\varnothing$ and $\mathbb{E}\{\cdot\}$ denote the imaginary unit, an empty set, and the statistical expectation, respectively.

\section{System description}\label{sec:system}
\begin{figure*}[t]
    \centering
    \includegraphics[width=1.2\columnwidth]{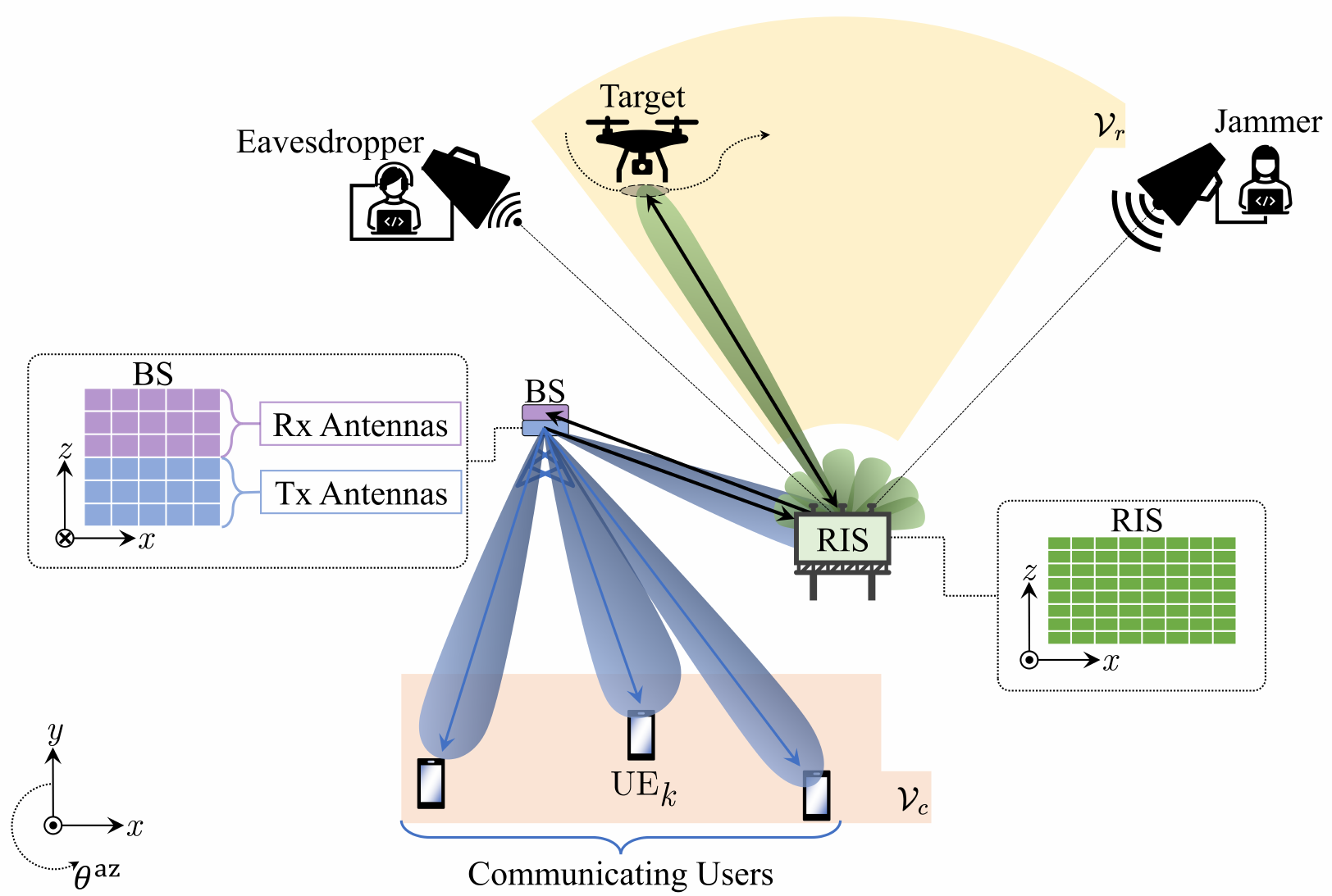}
    \caption{Considered architecture. The BS simultaneously serves multiple downlink ground users in the volume  $\mathcal{V}_{c}$ and, with the help of a passive RIS, verifies the existence of an airborne target in the volume $\mathcal{V}_{r}$. Eavesdroppers and/or jammers may also be present in the field of view of the RIS.}
    \label{fig:sm}
\end{figure*}
Consider a full-duplex BS equipped with two planar arrays, as shown in Fig.~\ref{fig:sm}; one array has $D_{\rm tx}$ antennas and is used for transmission, while the other one has $D_{\rm rx}$ antennas and is used for reception. The BS adopts an OFDM transmission format with a central carrier frequency $f_{o}$, a number of subcarriers $N_{o}$, a subcarrier spacing $W_{o}$, and a cyclic prefix of duration $T_{o}$, whereby the duration of each OFDM symbol is $T_{\rm sym}=1/W_{o}+T_{o}$. A subset of $N_{\rm sub}$ equally-spaced subcarriers is employed to simultaneously serve $K$ single-antenna ground users in the volume $\mathcal{V}_{c}$ and to detect a prospective airborne target in the volume  $\mathcal{V}_{r}$, with $\mathcal{V}_{c}\cap \mathcal{V}_{r} = \varnothing$; the corresponding subcarrier frequencies are $f_{1},\ldots,f_{N_{\rm sub}}$, with $f_{q}-f_{q-1}=W_{{\rm sub}}$ for $q=2,\ldots,N_{\rm sub}$. The operations on other subcarriers are not considered here. The radar function is implemented via a passive RIS, which is equipped with $D_{\rm ris}$ reflective elements arranged into a rectangular planar array. The RIS scans a volume not directly observable by the BS, thus resulting into a non-line-of-sight mono-static radar architecture~\cite{Buzzi-2022}. For future reference, we denote by $\mathcal{H}_{0}$ and $\mathcal{H}_{1}$ the hypotheses that no target and one target is present in the inspected volume $\mathcal{V}_{r}$, respectively. 

We assume that the users are not in the field-of-view of the RIS due to the specific system geometry and/or the presence of a blockage.\footnote{The following developments can be adapted to a different scenario where the RIS also supports the communication function. For example, if the RIS is partitioned into two sub-arrays~\cite{10360201}, one sub-array can be dedicated to the sensing function and the other one to the communication function.}
Also, we make a narrowband assumption on each subcarrier, whereby $(\Delta_{\rm tx}+\Delta_{\rm rx}+\Delta_{\rm ris})\ll c_{o}/W_{o}$, where $\Delta_{\rm tx}$,  $\Delta_{\rm rx}$, and $\Delta_{\rm ris}$ are the largest distance among any element pair in the transmit array, receive array, and RIS, respectively, and $c_{o}$ is the speed of light~\cite{Van-Trees-IV}. Finally, let $\tau_{o}$ be the maximum delay over any BS-user or BS-RIS-target-RIS-BS path; we assume $\tau_{o}\leq T_{o}$, so that no interference among subsequent OFDM symbols is present at the user and radar receivers. 

The volume $\mathcal{V}_{r}$  is in the  RIS far-field and, when observed from a Cartesian reference system at the center of gravity of the RIS, spans the range, azimuth, and elevation intervals $[R_{\min},R_{\max}]$, $[\theta_{\min}^{\rm{az}},\theta_{\max}^{\rm{az}}]$, and  $[\theta_{\min}^{\rm{el}},\theta_{\max}^{\rm{el}}]$, respectively; the far-field assumption implies that $R_{\min}\geq 2\Delta_{\rm ris}^2/ \lambda_{-}$, where $\lambda_{-}=c_{o}/(f_{o}- 0.5N_{o}W_{o})$ is the minimum wavelength~\cite{book-Stutzman03}. Let $\bm{\theta}= [\theta^{\rm az}\;\theta^{\rm el}]\transp$ be the angular direction specified by azimuth and elevation angles $\theta^{\rm az}$ and $\theta^{\rm el}$, respectively; then, we denote by $\bm{t}_{q}(\bm{\theta})$ the far-field steering vector of the RIS towards $\bm{\theta}$ on the $q$-th subcarrier, for $q=1,\ldots,N_{\rm sub}$.

\begin{figure*}[t!]
    \centering
    \includegraphics[width=0.9\linewidth]{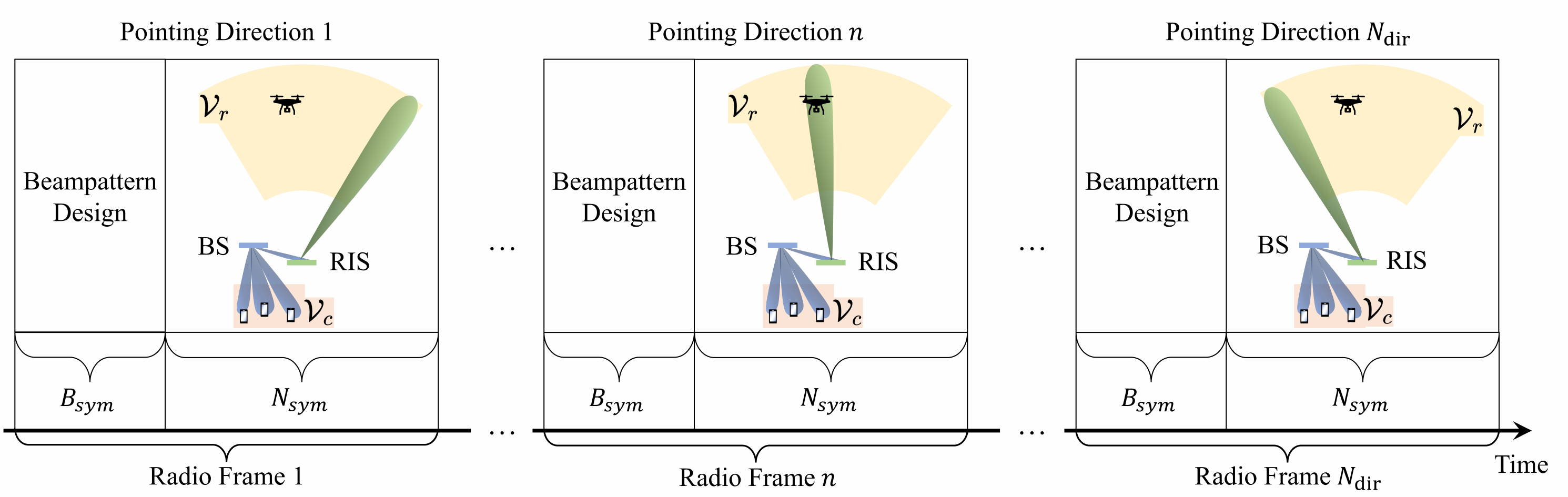}
    \caption{Scanning policy for the considered ISAC scenario.}
    \label{fig:sm_Scan}
\end{figure*}

The volume $\mathcal{V}_{r}$ is divided into $N_{\rm dir}$ conical subvolumes. We denote by $\bar{\bm{\theta}}_{i}=[\bar\theta_{i}^{\rm az}\;\bar\theta_{i}^{\rm el}]\transp$ the  nominal pointing direction corresponding to the $i$-th subvolume and by $\mathcal{G}_{\rm dir}=\{\bar{\bm{\theta}}_{1},\ldots,\bar{\bm{\theta}}_{N_{\rm dir}}\}$ the set of all such directions. As illustrated in Fig.~\ref{fig:sm_Scan}, the RIS-assisted radar sequentially illuminates each subvolume for a Coherent Processing Interval (CPI) containing $N_{\rm sym}$ OFDM symbols, say $T_{\rm cpi}=N_{\rm sym}T_{\rm sym}$, and it remains inactive for $B_{\rm sym}$ OFDM symbols between consecutive illuminations. The idle time is employed to elaborate the echoes received in the previous subvolume and reconfigure the RIS. Accordingly, the volume $\mathcal{V}_{r}$ is scanned in a time interval $T_{\rm scan}=N_{\rm dir}T_{\rm dir}$, where $T_{\rm dir}=(N_{\rm sym}+B_{\rm sym})T_{\rm sym}$.

In the following, we introduce the signal transmitted by the BS in a given CPI and the corresponding signals received by the radar and the users. To simplify exposition, we considered a CPI spanning the OFDM symbols indexed by $1,\ldots,N_{\rm sym}$.

\subsection{Transmit signal} 
In the considered CPI, let $\bm{f}_{q,k}$ and $\bm{f}_{q,r}$ be the unit-norm beamformers employed on the $q$-th subcarrier to send the sequence of data symbols $x_{q,k}(1),\ldots,x_{q,k}(N_{\rm sym})$ towards the $k$-th user and the sequence of dummy symbols $x_{q,r}(1),\ldots,x_{q,r}(N_{\rm sym})$ dedicated to the radar function, respectively, for $q=1,\ldots,N_{\rm sub}$ and $k=1,\ldots,K$; in the following, we refer to  $\{\bm{f}_{q,1},\ldots,\bm{f}_{q,K}\}_{q=1}^{N_{\rm{sub}}}$ and $\{\bm{f}_{q,r}\}_{q=1}^{N_{\rm{sub}}}$ as the communication and radar beamformers, respectively. Then, the signal emitted by the BS on the $q$-th subcarrier  in the $n$-th symbol interval is
 \begin{equation}
 	\bm{s}_{q}(n)=\sqrt{\mathcal{P}}\bm{F}_{q}\bm{x}_{q}(n),
 \end{equation} 
for $q=1,\ldots,N_{\rm sub}$ and $n=1,\ldots,N_{\rm sym}$, where $\mathcal{P}$ is the power radiated on each subcarrier, $\bm{F}_{q}=[\bm{f}_{q,1} \, \cdots \,  \bm{f}_{q,K}\, \bm{f}_{q,r}]\in\mathbb{C}^{D_{\rm tx}\times (K+1)}$, and $\bm{x}_{q}(n)=[x_{q,1}(n) \, \cdots \, x_{q,K}(n)\, x_{q,r}(n)]\transp\in\mathbb{C}^{K+1}$. We assume that $\{\bm{x}_{q}(1),\ldots,\bm{x}_{q}(N_{\rm sym})\}_{q=1}^{N_{\rm{sub}}}$ are  independent and identically distributed random vectors with zero-mean and  $\mathbb{E}[\bm{x}_{q}(n)\bm{x}_{q}\herm(n)]=\diag\{\bm{\gamma}_{q}\}$, where
\begin{equation}
	\bm{\gamma}_{q}=[\gamma_{q,1}\,\cdots\, \gamma_{q,K}\, \gamma_{q,r}]\transp.
\end{equation}
The entries of $\bm{\gamma}_{q}$ are positive and sum to one, thus specifying the fraction of power assigned to each stream; in particular, $\gamma_{q,c}=\sum_{k=1}^{K}\gamma_{q,k}$ and $\gamma_{q,r}=1-\gamma_{q,c}$ are the fraction of power assigned to the communication and radar functions, respectively. In the following, we refer to $\{\bm{F}_{q}\}_{q=1}^{N_{\rm{sub}}}$ and $\{\bm{\gamma}_{q}\}_{q=1}^{N_{\rm{sub}}}$ as the BS precoding matrices and power vectors, respectively.

\subsection{Radar received signal} 
In the considered CPI, let $\bm{\omega}=[\omega_1\,\cdots\,\omega_{D_{\rm ris}}]\transp\in\Omega^{D_{\rm ris}}$ be the vector modeling the RIS response, with $\Omega=\{\chi\in\mathbb C: |\chi|=1\}$. Then,
the signal received by the radar on the $q$-th subcarrier in the $n$-th symbol interval  is
\begin{subequations}\label{eq:rx_signal}
\begin{align}
\bm{y}_{q,r}(n)&=\bm{z}_{q,r}(n), &\text{under }\mathcal{H}_{0}, \label{eq:rx_signal_H0}\\
\bm{y}_{q,r}(n)&=\alpha\e^{-\i 2\pi q W_{\rm sub} \tau} \e^{-\i 2\pi \nu nT_{\rm sym}}\notag\\
&\quad \times \bm{G}_{q,\rm rx}\diag\{\bm{\omega}\}\bm{t}_{q}(\bm{\phi})\notag \\
&\quad \times \bm{t}_{q}\transp(\bm{\phi})\diag\{\bm{\omega}\}\bm{G}_{q,\rm tx}\bm{s}_{q}(n)\notag\\
&\quad +\bm{z}_{q,r}(n), &\text{under }\mathcal{H}_{1}, \label{eq:rx_signal_H1}
\end{align}
\end{subequations}
for $q=1,\ldots,N_{\rm sub}$ and $n=1,\ldots,N_{\rm sym}$, where: $\bm{z}_{q,r}(n)\in\mathbb{C}^{D_{\rm rx}}$ is the noise, modeled as a circularly-symmetric Gaussian random vector with covariance matrix $\sigma^{2}_{r}\bm{I}_{D_{\rm rx}}$; $\bm{G}_{q,\rm tx}\in\mathbb{C}^{D_{\rm ris}\times D_{\rm tx}}$ and $\bm{G}_{q,\rm rx}\in\mathbb{C}^{D_{\rm rx}\times D_{\rm ris}}$ are the channel matrices between the BS transmitter and the RIS and between the RIS and the BS receiver, respectively; finally, $\alpha$, $\tau$, $\nu$, $\phi^{\rm az}$, and $\phi^{\rm el}$ are the complex amplitude, delay, Doppler shift, azimuth angle, and elevation angle of the target, respectively.

We assume that the channel matrices $\bm{G}_{q,\rm rx}$ and $\bm{G}_{q,\rm tx}$ remain constant over the CPI and are known to the BS; for example, they can be estimated in the radar idle periods.\footnote{The BS-to-RIS channel changes over long time scales when both BS and RIS are mounted on top of tall structures: this implies that an accurate estimate of the BS-to-RIS channel can be obtained in practice.}  In the considered CPI, we define the two-way (power) beampattern of the RIS-aided radar on the $q$-th subcarrier as
\begin{align}
	\text{BP}_{q}(\bm{\theta})	&= \frac{1}{\mathcal{P}}\mathbb{E}\Big[\big\|\bm{G}_{q,\rm rx}\diag\{\bm{\omega}\}\bm{t}_{q}(\bm{\theta})\notag \\
	&\quad \times \bm{t}_{q}\transp(\bm{\theta})\diag\{\bm{\omega}\}\bm{G}_{q,\rm tx}\bm{s}_{q}(n)\big\|^2\Big] \notag \\
	&= \underbrace{\big\|\bm{G}_{q,\rm rx}\diag\{\bm{t}_{q}(\bm{\theta})\}\bm{\omega}\big\|^2}_{\text{BP}_{q,{\rm rx}}(\bm{\theta})}\notag \\
	&\quad \times \underbrace{\big\|\bm{\omega}\transp \diag\{\bm{t}_{q}(\bm{\theta})\}\bm{G}_{q,\rm tx}\bm{F}_{q}\diag\{\bm{\gamma}_{q}\}^{\frac{1}{2}}\big\|^2}_{\text{BP}_{q,{\rm{tx}}}(\bm{\theta})},\label{BP_per_sub}
\end{align}
where $\text{BP}_{q,{\rm tx}}(\bm{\theta})$ and $\text{BP}_{q,{\rm rx}}(\bm{\theta})$ are the transmit and receive beampatterns, which accounts for the power transmitted towards $\bm{\theta}$ and the power received from $\bm{\theta}$, respectively. We also define the two-way, transmit, and receive beampatterns of the RIS-aided radar over the entire frequency band as
\begin{subequations}
\begin{align}
\text{BP}(\bm{\theta}) &= \sum_{q=1}^{N_{\rm sub}}\text{BP}_{q}(\bm{\theta}), \label{seq:BP}\\
\text{BP}_{\rm{tx}}(\bm{\theta}) &= \sum_{q=1}^{N_{\rm{sub}}}\text{BP}_{q,\text{tx}}(\bm{\theta}),\label{seq:BP_tx}\\
\text{BP}_{\rm{rx}}(\bm{\theta}) &= \sum_{q=1}^{N_{\rm{sub}}}\text{BP}_{q,\text{rx}}(\bm{\theta}),\label{seq:BP_rx}
\end{align}
\end{subequations}
respectively. Notice that the above beampatterns depend on the BS power vectors $\{\bm{\gamma}_{q}\}_{q=1}^{N_{\rm{sub}}}$,  the BS precoding matrices $\{\bm{F}_{q}\}_{q=1}^{N_{\rm{sub}}}$, and the RIS response $\bm{\omega}$ (more on this in Sec.~\ref{sec:system_design}). Finally, from~\eqref{eq:rx_signal_H1}, \eqref{BP_per_sub}, and~\eqref{seq:BP}, the signal-to-noise ratio (SNR) of the target under $\mathcal{H}_{1}$ is
\begin{equation}
\mathrm{SNR}_{r}=\frac{N_{\rm{sym}}\mathcal{P}\,\text{BP}(\bm{\phi})\,\mathbb{E}\{|\alpha|^2\}}{\sigma^{2}_{r}}.
\label{eq:SNR_r}
\end{equation}

\subsection{User received signal}
In the considered CPI, the signal received by the $k$-th user on the $q$-th subcarrier in the $n$-th symbol interval is
\begin{align}
y_{q,k}(n) &=\underbrace{\sqrt{\mathcal{P}}\bm{h}_{q,k}\herm \bm{f}_{q,k}x_{q,k}(n)}_{\text{signal of interest}}\notag \\
 &\quad+\underbrace{\sum_{{\bar{k}=1,\; \bar{k}\neq k}}^{K}\sqrt{\mathcal{P}}\bm{h}_{q,k}\herm \bm{f}_{q,\bar{k}}x_{q,\bar{k}}(n)}_{\text{multi-user interference}}\notag\\ &\quad+\underbrace{\sqrt{\mathcal{P}}\bm{h}_{q,k}\herm \bm{f}_{q,r}x_{q,r}(n)}_{\text{radar interference}}+\underbrace{z_{q,k}(n)}_{\text{noise}},\label{user-rx-signal}
\end{align}
for $k=1,\ldots,K$, $q=1,\ldots,N_{\rm sub}$, and $n=1,\ldots,N_{\rm sym}$, where $\bm{h}_{q,k}\in\mathbb{C}^{D_{\rm tx}}$ is the channel vector between the BS and the $k$-th user on the $q$-th subcarrier and $z_{q,k}(n)\in\mathbb{C}$ is the additive noise, modeled as a circularly-symmetric Gaussian random variable with variance $\sigma^{2}_{k}$. In the $q$-th subcarrier, let $\bm{H}_{q}= [\bm{h}_{q,1}\,\cdots\,\bm{h}_{q,K}]\in\mathbb{C}^{D_{\rm tx}\times K}$ be the downlink channel matrix.  We assume that $\bm{H}_{q}$ is full rank, constant over the CPI, and known to the BS: for example, this matrix can be estimated in the radar idle periods.\footnote{To simplify exposition, perfect CSI between BS and users is assumed in this study.  Indeed, designing the communication beamformers at the BS and assessing the impact of possible channel estimation errors on the communication performance is not main focus of this study.} Accordingly, the SINR of the $k$-th user on the $q$-th subcarrier is
\begin{equation}
\mathrm{SINR}_{q,k}\!\!=\!\!\frac{|\bm{h}_{q,k}\herm \bm{f}_{q,k}|^2\gamma_{q,k}}{\displaystyle \sum_{{\bar{k}=1,\; \bar{k}\neq k}}^{K}\!\!\!\!\!\!\! |\bm{h}_{q,k}\herm \bm{f}_{q,\bar{k}}|^2\gamma_{q,\bar{k}}\!+\!\!|\bm{h}_{q,k}\herm \bm{f}_{q,r}|^2\gamma_{q,r}\!+\!\frac{\sigma^{2}_{k}}{\mathcal{P}}},
\end{equation}
for $k=1,\ldots,K$ and $q=1,\ldots,N_{\rm sub}$.  Assuming Gaussian signaling, the number of bits that can be reliably transmitted to user $k$ on the $q$-th subcarrier within one OFDM symbol is $\log_{2}(1+\mathrm{SINR}_{q,k})$; hence, the achievable rate (in bits per OFDM symbol) for user $k$ is
\begin{equation}
\mathcal{R}_{k}=\sum_{q=1}^{N_{\rm{sub}}}\log_{2}(1+\mathrm{SINR}_{q,k}), 
\end{equation}
for $k=1,\ldots,K$. Finally, the user sum-rate is 
\begin{equation}\label{user-sum-rate}
\mathcal{R}=\sum_{k=1}^{K}\mathcal{R}_{k}.
\end{equation}
Notice that the rate of each user depends on the BS power vectors $\{\bm{\gamma}_{q}\}_{q=1}^{N_{\rm{sub}}}$ and the BS precoding matrices $\{\bm{F}_{q}\}_{q=1}^{N_{\rm{sub}}}$ (more on this in Sec.~\ref{sec:system_design}).
 
\section{System design}\label{sec:system_design}

The BS power vectors $\{\bm{\gamma}_{q}\}_{q=1}^{N_{\rm{sub}}}$, the BS precoding matrices $\{\bm{F}_{q}\}_{q=1}^{N_{\rm{sub}}}$, the RIS response $\bm{\omega}$, and the implementation of the radar detector are under the control of the system engineer and can be optimized to obtain different tradeoffs between communication and radar performance. Consequently, we propose the following design strategy.
\begin{enumerate}
\item \label{step-1} The power split between the communication and radar functions on each subcarrier (i.e., the value of $\gamma_{q,c}=1-\gamma_{q,r}$ for $q=1,\ldots,N_{\rm{sub}}$) is kept fixed in all CPIs.

\item \label{step-2} In each CPI, the radar beamformer $\bm{f}_{q,r}$ is forced to belong to the orthogonal complement of the column-space of $\bm{H}_{q}$ to avoid any radar interference at the users, i.e., it is required that 
\begin{equation}\bm{H}_{q}\herm\bm{f}_{q,r}=\bm{0}_{K},\label{orth_constraint}
\end{equation}
for $q=1,\ldots,N_{\rm sub}$. Also, the fraction of power allocated to each user on the $q$-th subcarrier and the corresponding beamformers, namely, $\{\gamma_{q,k}\}_{k=1}^{K}$ and $\{\bm{f}_{q,k}\}_{k=1}^{K}$, are chosen to optimize the communication function, for $q=1,\ldots,N_{\rm sub}$: any of the design criteria already devised in the literature can be employed here:\footnote{Notice in passing that meaningful designs usually ensure that the vectors $\{\bm{f}_{q,k}\}_{k=1}^{K}$ belong to the column-space of $\bm{H}_{q}$, as any power sent outside such linear subspace will not be received by the users.} for example, see~\cite{zf_methods,Venturino-2015,10309244}.

\item \label{step-3} In each CPI, the radar beamformers $\{\bm{f}_{q,r}\}_{q=1}^{N_{\rm sub}}$ and the RIS response $\bm{\omega}$ are designed to steer the two-way beampattern of the RIS-aided radar toward the subvolume under inspection, while controlling its sidelobes in the presence of eavesdroppers, jammers, and other scattering objects, as described in Sec.~\ref{sec:beampattern}. 

\item \label{step-4} Finally, at the radar receiver, a multi-frame  detector is employed to jointly elaborates $N_{\rm scan}$ consecutive scans, so as to extend the time-on-target; to cope with the target motion in consecutive scans, a TBD-based processing in employed, as described in Sec.~\ref{sec:TBD}. 
\end{enumerate}

The proposed approach is quite flexible. In step~(\ref{step-1}), different physical resources can be granted to the communication and radar functions by varying $\gamma_{1,c},\ldots, \gamma_{N_{\rm{sub}},c}$; in step~(\ref{step-2}), different rates can be delivered to the users by resorting to different design criteria; in step~(\ref{step-3}), different scanning beams can be devised; in step~(\ref{step-4}), the target detection and estimation performance can be varied by increasing the number of processed scans. This latter point is attractive, since a larger complexity of the radar detector (i.e., a larger value of $N_{\rm scan}$) can be traded for a smaller value of $\gamma_{q,r}$ (and therefore a larger value of $\gamma_{q,c}$), while maintaining the same radar performance and improving the communication performance: an in-depth analysis will be provided in Sec.~\ref{sec:Numerical analysis}.

\subsection{Design of the radar beamformers and RIS response}\label{sec:beampattern}
We tackle here the problem of designing the radar beamformers $\{\bm f_{q,r}\}_{q=1}^{N\text{sub}}$ and the RIS response $\bm{\omega}$ to obtain a desired spatial response of the RIS-assisted radar transceiver in the considered CPI. The power sent to and received from the subvolume under inspection should be maximized to facilitate the discovery of a prospective target, while keeping under control the power sent to and received from other subvolumes to limit the false alarms caused by objects located outside the region of interest~\cite{Skolnik-book}. Moreover, if a malicious eavesdropper is present in the field of view of the RIS (see Fig.~\ref{fig:sm}), the power sent to the eavesdropper should be controlled in order to guarantee the security of the communication and/or realize a covert wireless system~\cite{Chen-2023}; similarly, if an intentional or unintentional jammer is present in the field of view of the RIS (see Fig.~\ref{fig:sm}), the power received from its direction should be limited to avoid blinding the radar~\cite{Richards-book}.

Let $\Theta_i$ and $\Theta_{i,{\rm sl}}$ be the finite discrete sets of angular directions spanning the $i$-th subvolume under inspection and the corresponding subvolume where the sidelobe level of the two-way beampattern has to be controlled, respectively. Similarly, let $\Theta_{\rm ev}$ and $\Theta_{\rm ja}$ be the finite discrete sets of angular directions where eavesdroppers and jammers may be present, respectively. Also, let $\bm U_q\in\mathbb C^{D_{\rm tx} \times (D_{\rm tx}-K)}$ be a matrix whose columns form an orthonormal basis of the null-space of $\bm H_q$. Then, the condition $\bm{H}_{q}\herm\bm{f}_{q,r}=\bm{0}_{K}$ implies that the radar beamformer $\bm f_{q,r}$ can be expanded as $\bm f_{q,r}= \bm U_q \tilde{\bm f}_{q,r}$, wherein $\tilde{\bm f}_{q,r} \in \mathbb C^{D_{\rm tx}-K}$ is under the designer's control, for $q=1,\ldots,N_{\rm{sub}}$. 
Then, the proposed design problem is formulated as
\begin{subequations}\label{eq:BP_problem_formulation_max}
  \begin{align}
    \max_{\substack{\bm{\omega}\in\Omega^{D_{\rm ris}}, \\ \{\tilde{\bm f}_{q,r}\}_{q=1}^{N_{\rm{sub}}}\\\in \mathbb C^{D_{\rm tx}-K}}} &\sum_{\bm \theta\in\Theta_i} \text{BP}\left(\bm{\theta};\bm{\omega}, \{\tilde{\bm f}_{q,r}\}_{q=1}^{N_{\rm{sub}}}\right), \\
    \text{s.t.} &\; \text{BP}  \left(\bm{\theta};\bm{\omega},\{\tilde{\bm f}_{q,r}\}_{q=1}^{N_{\rm{sub}}}\right) \leq \varepsilon_{\rm sl}, \forall \bm{\theta}\in\Theta_{i,{\rm sl}},\\
    &\; \text{BP}_{\rm{tx}} \!\left(\bm{\theta};\bm{\omega}, \{\tilde{\bm f}_{q,r}\}_{q=1}^{N_{\rm{sub}}}\right) \!\leq\! \varepsilon_{\rm ev} , \forall \bm{\theta} \in \Theta_{\rm ev},\\
    &\; \text{BP}_{\rm{rx}} \left(\bm{\theta};\bm{\omega}\right) \leq \varepsilon_{\rm ja} , \forall \bm{\theta} \in \Theta_{\rm ja} ,\notag\\
    &\; \| \bm U_q \tilde{\bm f}_{q,r} \|^2 =1, \; q=1,\ldots,N_{\rm sub},
  \end{align}
  \end{subequations}
where the dependence on the optimization variables $\bm{\omega}$ and $\{\tilde{\bm f}_{q,r}\}_{q=1}^{N_{\rm{sub}}}$ has been made explicit in the two-way, transmit, and receive beampatterns; in Problem~\eqref{eq:BP_problem_formulation_max}, $\varepsilon_{\rm sl}, \varepsilon_{\rm ev}, \varepsilon_{\rm ja}$ are upper bounds to the two-way beampattern in the sidelobe region, to the transmit beampattern in the eavesdropper region, and to the receive beampattern to the jammer region, while the last constraint forces the radar beamformers to be unit-modulus.

Problem~\eqref{eq:BP_problem_formulation_max} is non-convex and mixed-integer. Upon resorting to the penalty method, the constrained optimization problem is replaced by a series of one or more unconstrained problems obtained by penalizing the objective function with an extra cost (penalty function) that \emph{discourages} the violations of the constraints. In particular, we propose to approximate Problem~\eqref{eq:BP_problem_formulation_max} with Problem~\eqref{eq:BP_problem_formulation_penalty} reported at the top of the next page, where $\beta>0$ is a conveniently large number and $g(\chi)= (\max\{0,\chi\})^2$ is the standard quadratic penalty function~\cite{Bertsekas_1999}.\footnote{Notice that we can get arbitrary close to the optimal objective value of~\eqref{eq:BP_problem_formulation_max} by solving~\eqref{eq:BP_problem_formulation_penalty} with $\beta$ sufficiently large~\cite[Th.~9.2.2]{Bertsekas_1999}.}
\begin{figure*}
\begin{subequations}\label{eq:BP_problem_formulation_penalty}
\begin{align}
\max_{\substack{ \bm{\omega} \in \Omega^{D_{\rm ris}}, \\ \{\tilde{\bm f}_{q,r}\}_{q=1}^{N_{\rm{sub}}}\in \mathbb C^{D_{\rm tx}-K}}} 
&\Biggl\{ \sum_{\bm \theta\in\Theta_i} \text{BP}(\bm\theta;\bm{\omega}, \{ \tilde{\bm f}_{q,r}\}_{q=1}^{N_{\rm{sub}}}) -\beta \Biggl[\sum_{\bm \theta \in \Theta_{i,{\rm sl}}} g\Bigl( \text{BP} (\bm\theta;\bm{\omega}, \{\tilde{\bm f}_{q,r}\}_{q=1}^{N_{\rm{sub}}} )-\varepsilon_{\rm sl} \Bigr)\Biggr] \notag\\
&-\beta \Biggl[\sum_{\bm \theta \in \Theta_{\rm ev}} g \Bigl( \text{BP}_{\rm{tx}} (\bm\theta;\bm{\omega}, \{\tilde{\bm f}_{q,r}\}_{q=1}^{N_{\rm{sub}}}) - \varepsilon_{\rm ev}\Bigr)\Biggr] -\beta \Biggl[\sum_{\bm \theta \in \Theta_{\rm ja}} g \Bigl( \text{BP}_{\rm{rx}} (\bm\theta;\bm{\omega}) - \varepsilon_{\rm ja}\Bigr)\Biggr]  \Biggr\}, \\
\text{s.t.} &\; \| \bm U_q \tilde{\bm f}_{q,r} \|^2 =1, \; q=1,\ldots,N_{\rm sub}.
\end{align}
\end{subequations}
\noindent\rule{\textwidth}{0.4pt}
\end{figure*}

Problem~\eqref{eq:BP_problem_formulation_penalty} is still quite complex, and we propose to find a suboptimal solution by resorting to the block-coordinate ascent method~\cite{Bertsekas_1999}: starting from a feasible point, the objective function is maximized with respect to each of the block of variables, taken in cyclic order, while keeping the other ones fixed at their previous values. The block variables are $\omega_1$, $\omega_2$,\ldots, $\omega_{D_{\rm ris}}$, and $\{\tilde{\bm f}_{q,r}\}_{q=1}^{N_{\rm{sub}}}$. Therefore, we are faced with $D_{\rm ris}+1$ reduced-complexity sub-problems: $D_{\rm ris}$ sub-problems for the maximization over each RIS phase and one sub-problem for the maximization over the radar beamformers.


As to the RIS phase $\omega_d$, a suboptimal solution can be found by resorting to the projected gradient algorithm. In particular, the gradient of the objective function of Problem~\eqref{eq:BP_problem_formulation_penalty} is  reported in~\eqref{eq:gradient_ris} at the top of this page, where $g'(\chi)=2 \max\{0,\chi\}$ is the derivative of $g(\chi)$, and
\begin{figure*}
\begin{align}
 \bm \nabla_{\!\!\bm{\omega}} & = 2 \sum_{\bm \theta\in\Theta_i} \sum_{q=1}^{N_{\rm sub}} \Bigl( \bm \omega\herm \bm B_q(\bm \theta;\bm\omega) \bm \omega \bm A_q(\bm \theta;\bm\omega) \bm \omega  + \bm \omega\herm \bm A_q(\bm \theta;\bm\omega)\bm \omega \bm B_q(\bm \theta;\bm\omega) \bm \omega \Bigr) \notag\\
 &\quad -2 \beta \Bigg[\sum_{\bm \theta \in \Theta_{i,{\rm sl}}} \!\!g'\Bigl( \text{BP} (\bm\theta;\bm{\omega}, \{\tilde{\bm f}_{q,r}\}_{q=1}^{N_{\rm{sub}}}) -\varepsilon_{\rm sl} \Bigr)\!\! \sum_{q=1}^{N_{\rm sub}}\! \Bigl( \bm \omega\herm \bm B_q(\bm \theta;\bm\omega) \bm \omega \bm A_q(\bm \theta;\bm\omega) \bm \omega + \bm \omega\herm \bm A_q(\bm \theta;\bm\omega)\bm \omega \bm B_q(\bm \theta;\bm\omega) \bm \omega \Bigr) \Bigg]\notag\\
 &\quad -2 \beta \Bigg[\sum_{\bm \theta \in \Theta_{\rm ev}} \!\!g' \Bigl( \text{BP}_{\rm{tx}} (\bm\theta;\bm{\omega}, \{\tilde{\bm f}_{q,r}\}_{q=1}^{N_{\rm{sub}}}) - \varepsilon_{\rm ev}\Bigr) \!\!\sum_{q=1}^{N_{\rm sub}}\!\! \bm B_q(\bm \theta;\bm\omega) \bm \omega\Bigg] \!\!-\!2 \beta \Bigg[\sum_{\bm \theta \in \Theta_{\rm ja}} \!\!g' \Bigl( \text{BP}_{\rm{rx}} (\bm\theta;\bm{\omega}) - \varepsilon_{\rm ja}\Bigr) \!\!\sum_{q=1}^{N_{\rm sub}}\!\! \bm A_q(\bm \theta;\bm\omega) \bm \omega\Bigg]\!, \label{eq:gradient_ris} 
\end{align}
\noindent\rule{\textwidth}{0.4pt}
\end{figure*}

\begin{subequations}\label{eq:A_q_B_q}
 \begin{align}
   \bm A_q(\bm \theta;\bm\omega) &= \diag\{\bm{t}_{q}^*(\bm{\theta})\} \bm{G}_{q,\rm rx}\herm \bm{G}_{q,\rm rx} \diag\{\bm{t}_{q}(\bm{\theta})\},\\
   \bm B_q (\bm \theta;\bm\omega) &= \diag\{\bm{t}_{q}^*(\bm{\theta})\} \bm{G}_{q,\rm tx}^* \bm{F}_{q}^* \diag\{\bm{\gamma}_{q}\} \notag \\
   &\quad \times \bm{F}_{q}\transp \bm{G}_{q,\rm tx}\transp \diag\{\bm{t}_{q}(\bm{\theta})\}.
 \end{align}%
\end{subequations}
Then, the $i$-th iteration of the algorithm is 
\begin{equation}\label{eq:grad_ascent_omega}
 \bm \omega^{(i)}= \Pi\left( \bm \omega^{(i-1)} + \lambda^{(i)} \bm \nabla_{\!\!\bm{\omega}}^{(i-1)} \right),
\end{equation}
where $\lambda^{(i)}$ is the step size at the $i$-th iteration, which can be computed, for example, though a backtracking Armijo line search~\cite{Nocedal_2006}, and $\Pi(\, \cdot\,)$ is the projection on $\Omega^{D_{\rm ris}}$, i.e., $\Pi(\bm \omega) = \hat{\bm \omega}$, with $\hat \omega_d=\omega_d/|\omega_d|$, for\footnote{If $\Omega$ is a finite discrete set, then the corresponding projection is given by  $\hat \omega_d=\argmin_{\chi\in\Omega} |\chi -\omega_d|$, for $d=1,\ldots,D_{\rm ris}$. The complex unit circle is a good approximation when the number of the RIS phase states is greater than or equal to 8~\cite{9223720}.} $d=1,\ldots,D_{\rm ris}$.

As to the radar beamformers $\{\tilde{\bm f}_{q,r}\}_{q=1}^{N_{\rm{sub}}}$, a suboptimal solution can be found by resorting again to the projected gradient method. Specifically, let $\bm \nabla_{q,r}$ be the vector of partial derivative of the objective function in Problem~\eqref{eq:BP_problem_formulation_penalty} with respect to $\tilde{\bm f}_{q,r}$,  that is reported in~\eqref{eq:Fr_part_derivative} at the top of the next page, where
\begin{figure*}
\begin{align}
 \bm \nabla_{q,r} &= 2\sum_{\bm \theta\in\Theta_i} \text{BP}_{q,\text{rx}}(\bm \theta;\bm \omega) \bm b_q (\bm \theta;\bm \omega) \bm b_q\herm(\bm \theta;\bm \omega)  \tilde{\bm f}_{q,r} \notag\\
 & \quad - 2\beta \Bigg[ \sum_{\bm \theta \in \Theta_{i,{\rm sl}}} g' \Bigl(\text{BP} (\bm\theta;\bm{\omega}, \{\tilde{\bm f}_{q,r}\}_{q=1}^{N_{\rm{sub}}}) -\varepsilon_{\rm sl}  \Bigr)  \text{BP}_{q,\text{rx}}(\bm \theta;\bm \omega) \bm b_q (\bm \theta;\bm \omega) \bm b_q\herm(\bm \theta;\bm \omega)  \tilde{\bm f}_{q,r}\Bigg]\notag \\
 & \quad - 2\beta \Bigg[ \sum_{\bm \theta \in \Theta_{\rm ev}} g' \Bigl(\text{BP}_{\rm{tx}}(\bm\theta;\bm{\omega}, \{\tilde{\bm f}_{q,r}\}_{q=1}^{N_{\rm{sub}}}) -\varepsilon_{\rm ev} \Bigr) \bm b_q (\bm \theta;\bm \omega) \bm b_q\herm(\bm \theta;\bm \omega)  \tilde{\bm f}_{q,r}\Bigg],\label{eq:Fr_part_derivative}
\end{align}
\noindent\rule{\textwidth}{0.4pt}
\end{figure*}

\begin{equation}\label{eq:b_q}
\bm b_q(\bm \theta;\bm \omega)\!=\!\sqrt{\gamma_{q,r}} \bm U_q\herm \bm{G}_{q,\rm tx}\herm \diag\{\bm{t}_{q}^*(\bm{\theta})\} \bm{\omega}^* \in \mathbb C^{D_{\rm{tx}}-K}\!\!\!.
\end{equation}
Then, the $i$-th iteration of the algorithm is 
\begin{equation}\label{eq:grad_ascent}
 \tilde{\bm f}_{q,r}^{(i)}=\frac{\tilde{\bm f}_{q,r}^{(i-1)} + \lambda^{(i)} \bm \nabla_{q,r}^{(i-1)}}{\bigl\Vert \bm U_q \bigl( \tilde{\bm f}_{q,r}^{(i-1)} + \lambda^{(i)} \bm \nabla_{q,r}^{(i-1)} \bigr)\bigr\Vert}, 
\end{equation}
for $q=1,\ldots,N_{\rm sub}$, where $\lambda^{(i)}$ is the step size at the $i$-th iteration; notice here that the normalization implements the projection on the feasible search set, thus ensuring that $\bm f_{q,r}= \bm U_q \tilde{\bm f}_{q,r}$ is unit-modulus.

Alg.~\ref{alg:BP_design} summarizes the proposed block-coordinate ascent routine. Since the projected gradient method starts from the point found at the previous iteration, the objective function is non-decreasing in successive iterations, and Alg.~\ref{alg:BP_design} converges. However, since Problem~\eqref{eq:BP_problem_formulation_penalty} is not convex and since the feasible search set cannot be expressed as the Cartesian product of closed convex sets, there is no guarantee that a global maximum is reached.

\begin{algorithm}[t]
  \caption{Suboptimal solution to Problem~\eqref{eq:BP_problem_formulation_penalty}} \label{alg:BP_design}
  \begin{algorithmic}[1]
    \STATE initialize $\bm{\omega}$ and $\{\tilde{\bm f}_{q,r}\}_{q=1}^{N_{\rm sub}}$
    \REPEAT
      \REPEAT
        \STATE compute $\bm \nabla_{\!\!\bm{\omega}}$ in~\eqref{eq:gradient_ris}
        \STATE set the step-size $\lambda$
        \STATE $\bm \omega= \bm \omega + \lambda \bm \nabla_{\!\!\bm{\omega}}$
        \STATE $\omega_d=\omega_d/|\omega_d|$, $\forall d$ 
      \UNTIL convergence
      \REPEAT
        \STATE compute $\bm \nabla_{q,r}$ in~\eqref{eq:Fr_part_derivative}, $\forall q$
        \STATE set the step-size $\lambda$
        \STATE $\tilde{\bm f}_{q,r}= \tilde{\bm f}_{q,r} + \lambda \bm \nabla_{q,r}$, $\forall q$
        \STATE $\tilde{\bm f}_{q,r}= \tilde{\bm f}_{q,r}/ \Vert\bm U_q \tilde{\bm f}_{q,r} \Vert$, $\forall q$
      \UNTIL convergence
    \UNTIL convergence
  \RETURN $\bm{\omega}$ and $\{\tilde{\bm f}_{q,r}\}_{q=1}^{N_{\rm sub}}$
  \end{algorithmic}
\end{algorithm}

\subsection{Design of the radar detector}\label{sec:TBD}
The measurements collected in each CPI are processed through a bank of correlators, each matched to a delay, say $\tau$, and a Doppler shift, say $\nu$, taken from the uniformly-spaced grids $\mathcal{G}_{\rm del}=\{\bar{\tau}_1,\ldots,\bar{\tau}_{N_{\rm del}}\}$ and $\mathcal{G}_{\rm dop}=\{\bar{\nu}_1,\ldots,\bar{\nu}_{N_{\rm dop}}\}$, respectively; each correlator is followed by a square law detector and a noise variance normalization. 
Following~\cite{Richards-book}, the output statistic is
\begin{align}\label{eq:filter_output}
A(\bm{\theta},\tau,\nu) =\frac{1}{\sigma_{r}^{2}}\left|\sum_{q=1}^{N_{\rm sub}}\sum_{n=1}^{N_{\rm sym}} \bm{u}_{q}\herm(n;\bm{\theta},\tau,\nu) \bm{y}_{q,r}(n)\right|^2
\end{align}
where 
\begin{align}
\bm{u}_{q}(n;\bm{\theta},\tau,\nu)&=w_{q}(n)\e^{-\i 2\pi q W_{\rm sub} \tau}\e^{-\i 2\pi \nu nT_{\rm sym}} \notag \\
&\quad \times\frac{\bm{G}_{q,\rm rx}\diag\{\bm{\omega}\}\bm{t}_{q}(\bm{\theta})}{\|\bm{G}_{q,\rm rx}\diag\{\bm{\omega}\}\bm{t}_{q}(\bm{\theta})\|}\notag \\
&\quad \times \frac{\bm{t}_{q}\transp(\bm{\theta})\diag\{\bm{\omega}\}\bm{G}_{q,\rm tx}\bm{s}_{q}(n)}{|\bm{t}_{q}\transp(\bm{\theta})\diag\{\bm{\omega}\}\bm{G}_{q,\rm tx}\bm{s}_{q}(n)|},\label{eq:filter}
\end{align} 
for $q=1,\ldots,N_{\rm sub}$ and  $n=1,\ldots,N_{\rm sym}$, are the tap amplitudes of the correlator matched to the triplet $(\bm{\theta},\tau,\nu)\in \mathcal{G}_{\rm dir}\times \mathcal{G}_{\rm del}\times \mathcal{G}_{\rm dop}$ and $\{w_{q}(n)\}$ are positive weights normalized to have $\sum_{q=1}^{N_{\rm sub}}\sum_{n=1}^{N_{\rm sym}}w^2_{q}(n)=1$. In~\eqref{eq:filter}, the knowledge of the transmitted signal is exploited to combine the measurements (and coherently integrate the energy) taken in different subcarriers and symbol intervals; also, the adopted normalization ensures that $\|\bm{u}_{q}(n;\bm{\theta},\tau,\nu)\|=1$; finally, the weights $\{w_{q}(n)\}$ can be chosen to obtain lower sidelobes in the range and Doppler domains at the price of a wider and lower mainlobe.

In the $\ell$-th scan, the statistics in~\eqref{eq:filter_output} are collected into an array $\bm{A}_{\ell}$ of size $N_{\rm dir} \times N_{\rm del}\times  N_{\rm dop}$, wherein the entry $(i,j,d)$ corresponds to the pointing direction $\bar{\bm{\theta}}_{i}$, the delay $\bar{\tau}_{j}$, and the Doppler shift $\bar{\nu}_{d}$. The raw data in $\bm{A}_{\ell}$ is then sent to the \emph{detector and plot-extractor}, which produces a list of measurements called \emph{plots} in the radar jargon. This block may involve many operations, but, at a high level, it can be modeled as a multi-peak detector, where a plot is formed for each peak in $\bm{A}_{\ell}$ exceeding a given threshold $\eta_{\rm plot}$. Here, a plot is defined as a 5-dimensional row vector containing the value of the test statistic in~\eqref{eq:filter_output}, the position (range, azimuth, and elevation) of the peak, and the time at which the subvolume containing the detected peak has been illuminated. The plots are organized in the plot-list $\bm S_\ell$, which is a 5-column matrix with the extracted plots as rows; if no plot is formed, $\bm S_\ell =\emptyset$.

\begin{figure}[t]
\centering
\centerline{\includegraphics[width=\columnwidth]{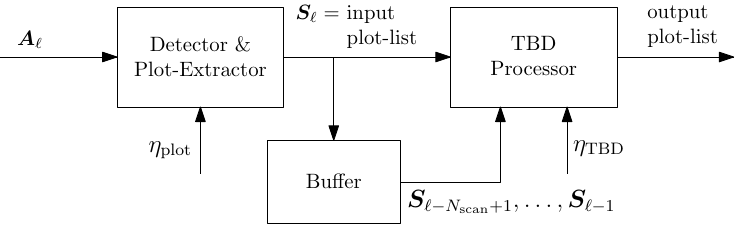}}
\caption{Block diagram of the considered radar detector.}
\label{fig:TBD_scheme}
\end{figure}

In a standard radar detector, the decision on the target presence is made on the basis of the current scan only. However, this approach shows severe limitations when the signal amplitude is weak compared to the background noise, which can be the case for non-line-of-sight mono-static radar architectures. To overcome this limitation, we propose here to make a decision on the target presence after integrating its signal returns over multiple consecutive scans. This is challenging in the presence of target motion, and TBD techniques\footnote{This terminology is used to mark the difference with respect to the classical approach, where the target is first detected and then tracked over multiple scans; with the TBD approach, instead, a number of scans are jointly processed with no (or little) thresholding taking place, and an estimated trajectory is returned at the same time as a detection is declared.} are required to properly integrate the echoes over consecutive scans~\cite{Pulford_2010}.Many approaches have been proposed in the literature. Here we resort to the multi-frame detector proposed in~\cite{Grossi-2013a, Grossi-2013b}, that has also been validated with real data in~\cite{Aprile-2016}. The detector adopts a computationally efficient dynamic programming algorithm to compute the required statistics. The considered TBD scheme is reported in Fig~\ref{fig:TBD_scheme}. The first stage contains the detector and plot-extractor, wherein the threshold is lowered to obtain a richer set of plots: this allows to preserve the measurements from dim targets at the price of increasing the number of false alarms. The second stage is the TBD processor, which exploits the space-time correlation among the plots in the current scan and those in the previous $N_{\rm scan}-1$ scans to confirm or delete them; the final decision on the target presence in the current scan is then made by comparing a convenient scoring metric against a threshold that is adjusted to have the desired probability of false alarm ($P_{\rm fa}$).

Next, we provide a short description of the TBD processor, while more details can be found in~\cite{Grossi-2013a, Grossi-2013b, Aprile-2016}. To simplify presentation, we assume that the current scan is $\ell=N_{\rm scan}$, so that the plot-lists taken in the scans $1,\ldots,N_{\rm scan}$ are  processed. The trajectory of a prospective target is specified by an $N_{\rm scan}$-dimensional vector, say $\bm \xi=[\xi_1 \; \cdots \; \xi_{N_{\rm scan}}]\transp$. Specifically, $\xi_i=k$ means that the target is observed at the $i$-th scan and the corresponding plot is the $k$-th row of $\bm S_i$; accordingly, $\bigl[S_i(k,2)\; S_i(k,3)\; S_i(k,4)\bigr]\transp$ is the target position at time $S_i(k,5)$, and $S_i(k,1)$ is the statistic resulting from~\eqref{eq:filter_output}. Instead, $\xi_i=0$ means that the target is not observed at the $i$-th scan. The metric used to evaluate the trajectory index by $\bm \xi$ is
\begin{equation}
 \mathcal T_{\bm \xi} = \sum_{\substack{i=1,\; \xi_i\neq 0}}^{N_{\rm scan}} S_i(\xi_i,1),
\end{equation}
which accounts for the overall energy back-scattered by a target moving along the trajectory indexed by $\bm \xi$. The decision on the target presence is finally made as
\begin{equation}
 \max_{\bm \xi\in \Xi} \mathcal T_{\bm \xi} \test \eta_{\rm TBD},
\end{equation}
where $\eta_{\rm TBD}$ is the detection threshold used in the TBD processor, and $\Xi$ is the set containing the vectors indexing all trajectories ending in a plot at scan $N_{\rm scan}$ and satisfying the required physical constraints on the target motion, such as maximum speed and acceleration. If a target is declared, the associated trajectory is indexed by $\argmax_{\bm \xi \in \Xi} \mathcal T_{\bm \xi}$.

The complexity of this TBD processor is linear in the number of integrated scans, $N_{\rm{scan}}$, and quadratic in the number of plots per scan, i.e, the number of rows in $\bm{S}_{\ell}$ for $\ell = 1, \ldots, N_{\rm{scan}}$, which makes its real time implementation feasible: a detailed discussion on this can be found in~\cite{Grossi-2013a,Grossi-2013b}. Also, it is worth mentioning that the trajectory of the detected target can be \emph{smoothed} by using polynomial regression: this improves the accuracy of the range and azimuth measurements based on the estimated history and, as a by-product, may provide information about the velocity of the targets~\cite{Aprile-2016}. Finally, notice that this method can readily be generalized to the case where multiple targets can be present in the inspected volume  $\mathcal{V}_{r}$: data association problems arising when more plots at the current scan correlate with the same plot at some previous scan can be solved through a \emph{successive track cancellation} procedure, with a complexity at most linear in the number of detected targets~\cite{Grossi-2013a,Grossi-2013b}.

\section{Numerical analysis}\label{sec:Numerical analysis}

Here we report an example of application for the architecture in Fig.~\ref{fig:sm}: the considered setup and the corresponding performance are analyzed in Secs.~\ref{subsec_setting} and~\ref{subsec_num_res}, respectively.

\subsection{System setup}\label{subsec_setting}

\begin{figure}[t]
    \centering
    \includegraphics[width=\columnwidth]{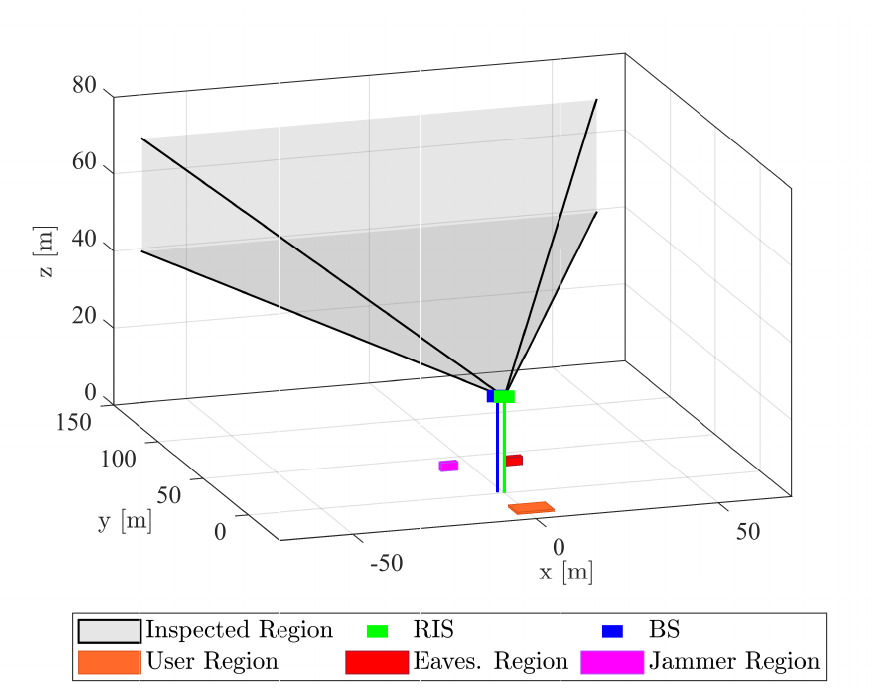}
    \caption{Considered system geometry.}
    \label{fig:map_3D}
\end{figure}

\begin{table}[t]
 \caption{System Parameters}\label{tab:parameters}
 \begin{center}
    \renewcommand{\arraystretch}{1.2} 
    \begin{tabular}{p{0.45\columnwidth}p{0.45\columnwidth}}
        \toprule
        \rowcolor{Gray}Carrier frequency & $f_{o}=3.5$~GHz\\	
        Spacing between the available\newline subcarriers & \mbox{}\newline $W_{o}=15$~kHz \\  
        \rowcolor{Gray} Number of available subcarriers & $N_{o}=3300$\\  
        Cyclic prefix duration & $T_{o}= 4.7623\upmu$s\\ 
        \rowcolor{Gray} OFDM symbol duration & $T_{\rm sym}=71.429 \upmu$s \\	
        Radio frame duration & $140  T_{\rm sym}=10$ ms\\
        \rowcolor{Gray} Radiated power per subcarrier & $\mathcal{P}= 4.803\upmu$W \\
        Available power & $N_{o}\mathcal{P}=15.85$ mW (12 dBm) \\ 
        \rowcolor{Gray} Spacing between employed\newline subcarriers & \mbox{}\newline $W_{\rm sub}=48 W_{o} = 720$~KHz \\  
        Number of employed subcarriers & $N_{\rm sub} = 32$\\
        \rowcolor{Gray} Number of symbols in a CPI & $N_{\rm sym} = 64$\\
        Number of guard symbols & $B_{\rm sym}=76$ \\
        \rowcolor{Gray} Number of RIS elements & $D_{\rm ris}= 8 \times 8=64$ \\ 
        Number of TX elements & $D_{\rm tx} =5 \times 3=15$ \\
        \rowcolor{Gray} Number of RX elements & $D_{\rm rx}= 5 \times 3=15$ \\
        RIS \& BS element spacing & $\lambda_{-}/2 =4.31$~cm \\
        \bottomrule    
    \end{tabular}
    \end{center}
\end{table}

The considered system geometry and parameters are reported in Fig.~\ref{fig:map_3D} and Table~\ref{tab:parameters}, respectively; in particular, typical 5G
signal specifications are considered here.

The BS and RIS are deployed in close proximity to reduce the multiplicative path loss along the indirect path, with all pairs of transmit and reflective elements in each other's far-field~\cite{Buzzi-2022}.  Also, the placement and orientation of the RIS is chosen to ensure a good illumination of the inspected volume $\mathcal{V}_{r}$. More specifically, the BS and RIS are located at $[-1.5\; 1.5\; 25]\transp$ and $[0\; 0\; 25]\transp$, respectively, with the Cartesian coordinates expressed in meters; their planar arrays are parallel to the ($x,z$)-plane and oriented towards the negative and positive $y$ axis, respectively (see also Fig.~\ref{fig:map_3D}).

The link between the $j$-th transmitting element and the $i$-th RIS element on the $q$-th subcarrier is modeled as
\begin{align}
\bigl[\bm{G}_{q,\rm tx}\bigr]_{i,j} =& \sqrt{\text{G}_{\rm{tx}}(\bm{\theta}_{i,j})\text{G}_{\rm{ris}}(\bm{\varphi}_{i,j})}\frac{\lambda_{q}}{4\pi d_{i,j}}\e^{-\i2\pi d_{i,j}f_{q}},
\end{align}
where $\text{G}_{\rm{tx}}(\bm{\theta})=\text{G}_{\rm{ris}}(\bm{\theta})=\pi \cos(\theta^{\rm{az}})\cos(\theta^{\rm{el}})$ is the element gain for the BS and the RIS, $\bm{\theta}_{i,j}$ is the angle of departure, $\bm{\varphi}_{i,j}$ is the angle of arrival, and $d_{i,j}$ distance. The backward link between the RIS and the BS is similarly modeled. 

The users and the BS are in each other's far-field, and a Ricean-distributed channel model is assumed. In particular, the downlink channel for the $k$-th user over the $q$-th subcarrier is
\begin{equation}
\bm{h}_{q,k} \!=\!\! \sqrt{\!\text{G}_{\rm{tx}}(\bm{\theta}_{k})\text{G}_{\rm{u}}(\bm{\varphi}_{k})}\frac{\lambda_{q}}{4\pi d_{k}}\e^{-\i2\pi d_{k}f_{q}}\bm{t}_{q}(\bm{\theta}_{k}) \!+ \tilde{\bm{h}}_{q,k},
\end{equation}
where $\text{G}_{\rm{u}}(\bm{\theta})=\pi \cos(\theta^{\rm{az}})\cos(\theta^{\rm{el}})$ is the antenna gain of the users, $\bm \theta_k$ is the angle of the departure, $\bm \varphi_k$ is the angle of arrival, $d_k$ is the distance, and $\tilde{\bm{h}}_{q,k} \sim \mathcal{CN}(\bm{0}_{D_{\rm tx}},\sigma_{\text{mp}}^2\bm{I}_{D_{\rm tx}})$ accounts for the random multipath component; the Ricean factor is set to $\text{G}_{\rm{tx}}(\bm{\theta}_{k})\text{G}_{\rm{u}} (\bm{\varphi}_{k})\lambda_{q}^2/(4\pi d_{k}\sigma_{\text{mp}})^2=20$~dB. Uniform power allocation is assumed (i.e., $\gamma_{q,k}=(1-\gamma_{q,r})/K$, for $k=1,\ldots,K$ and $q=\,\ldots,N_{\rm sub}$); also, a zero-forcing precoding is used at the BS~\cite{zf_methods}, whereby $\bm f_{q,k}$ is the $k$-th column of $\bm H_q^+$ normalized to have unit-norm. Finally, the noise variance of the users is $\sigma^{2}_{k}=F_{\rm noise} k_{\rm B} T_0 W_o=1.918\times 10^{-16}$~W, where $F_{\rm noise} =5$~dB is noise figure of the receiver, $k_{\rm B}$ is the Boltzmann constant, and $T_0=293$~K.

A uniform power allocation is also adopted for the radar function, whereby $\gamma_{q,r} = \gamma_{r}$ for $q=1,\ldots,N_{\rm sub}$. Swerling~I fluctuation is assumed for the target response, whereby $\alpha$ is modeled as a complex circularly-symmetric Gaussian random variable with variance dictated by the radar equation and equal to $\text{G}^2_{\rm ris} (\bm{\phi}) \sigma_{\text{RCS}} \lambda_{o}^2/ ((4\pi)^3d^4)$, where $\bm{\phi}$ and $d$ are the target's angle and distance from the RIS, and $\sigma_{\text{RCS}}$ is the target radar cross-section (RCS). The RCS is set to have a \emph{nominal} SNR in~\eqref{eq:SNR_r} equal to 27~dB at the current scan when all power is devoted to the radar function. Finally, the noise variance of the radar receiver is $\sigma^{2}_{r}=1.918\times 10^{-16}$~W.

\begin{figure}[t]
    \centering
    \includegraphics[width=\columnwidth]{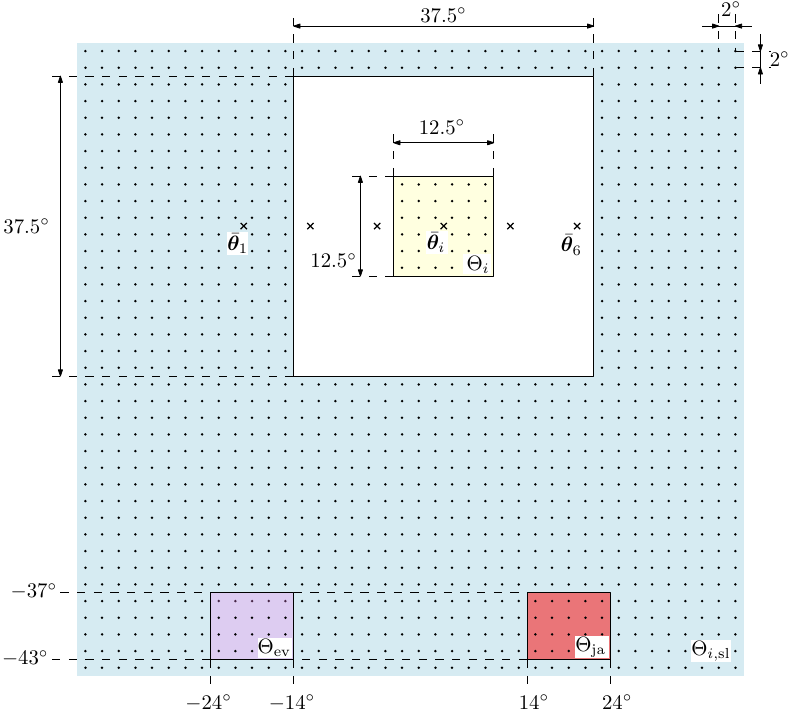}
    \caption{Jammer and eavesdropper regions, along with the sets $\Theta_i$, $\Theta_{i,{\rm sl}}$, $\Theta_{\rm ev}$, and $\Theta_{\rm ja}$ in Problem~\eqref{eq:BP_problem_formulation_max}.}
    \label{fig:regions}
\end{figure}

\begin{figure}[t]
    \centering
    \includegraphics[width=\columnwidth]{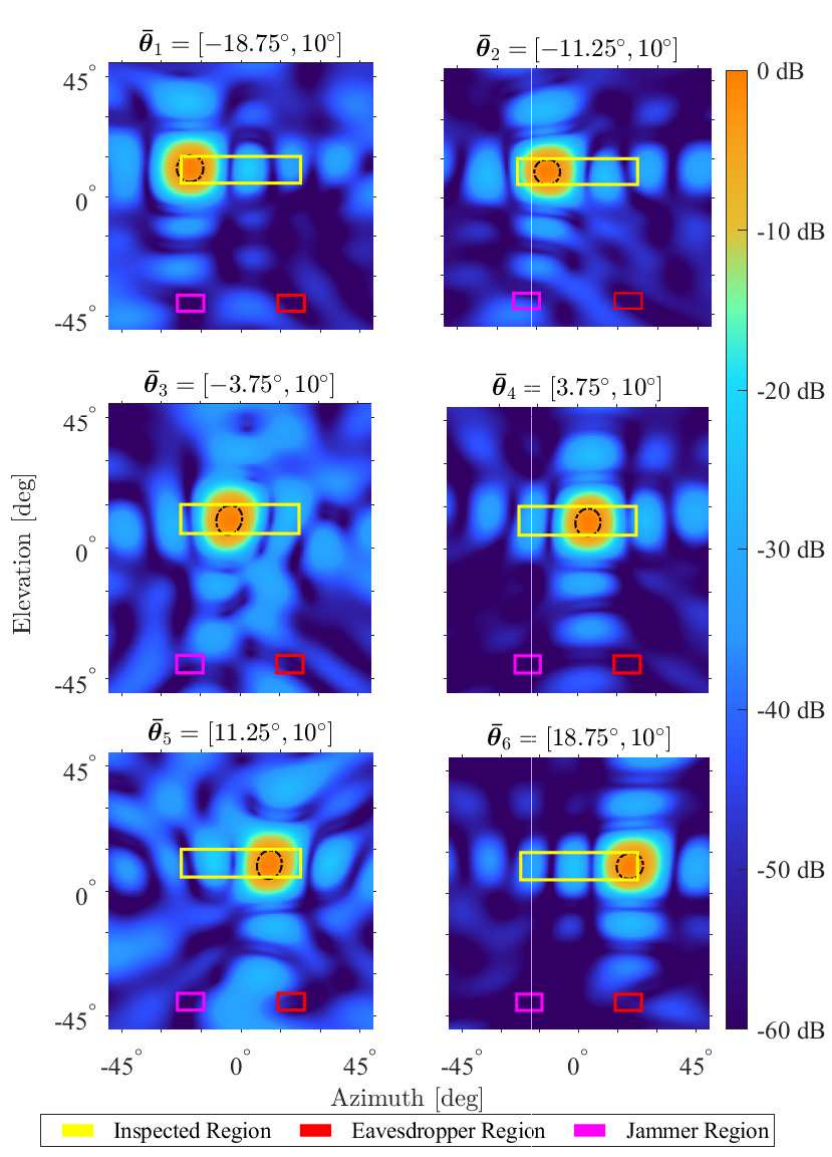}
    \caption{Normalized two-way beampattern for the considered pointing directions $\{\bar{\bm \theta}_i\}_{i=1}^6$. In each plot, the inspected, eavesdropper, and jammer regions are also shown, along with a dashed black contour line marking the 3-dB level of the mainlobe.}
    \label{fig:bp_map}
\end{figure}

With the considered system parameters, the (single-scan) resolution of the radar is $c_{o}/(2N_{\rm sub}W_{\rm sub}) =6.51$~m for the range and $\lambda_{o}/(2 N_{\rm sym}T_{\rm sym}) =9.37$~m/s for the radial velocity. Also, the maximum allowed range (limited by the cyclic prefix) is $c_{o}T_{o}/2 = 713.79$~m, while the non-ambiguous range and radial velocity are $c_{o}/(2W_{\rm sub})=208.19$~m and $\lambda_{o}/(4T_{\rm sym})=299.79$~m/s, respectively. The monitored volume  is scanned via $N_{\rm dir}=6$ illuminations with nominal pointing directions $\bar{\bm{\theta}}_{1}=[-18.75^\circ\;10^\circ]\transp$, $\bar{\bm{\theta}}_{2}=[-11.25^\circ\;10^\circ]\transp$, $\bar{\bm{\theta}}_{3}=[-3.75^\circ\;10^\circ]\transp$, $\bar{\bm{\theta}}_{4}=[3.75^\circ\;10^\circ]\transp$, $\bar{\bm{\theta}}_{5}=[11.25^\circ\;10^\circ]\transp$, and $\bar{\bm{\theta}}_{6}=[18.75^\circ\;10^\circ]\transp$. The jammer and eavesdropper regions are drawn in Fig.~\ref{fig:regions}, along with the sets $\Theta_i$, $\Theta_{i,{\rm sl}}$, $\Theta_{\rm ev}$, and $\Theta_{\rm ja}$ in Problem~\eqref{eq:BP_problem_formulation_max}. The two-way beampattern synthesized for $\varepsilon_{\rm sl}=\varepsilon_{\rm ev}=\varepsilon_{\rm ja}=10^{-8}$ is shown in Fig.~\ref{fig:bp_map}, where the dashed contour line marks the 3~dB beam that closely envelops each inspected subvolume under inspection. All the radar specifications are reported in Table~\ref{tab:radar_specs}.
 
\begin{table}[t]
 \caption{Radar Specifications}\label{tab:radar_specs}
 \begin{center}
    \renewcommand{\arraystretch}{1.2} 
    \begin{tabular}{p{0.45\columnwidth}p{0.45\columnwidth}}
        \toprule
        \rowcolor{Gray} Range resolution & $c_{o}/(2N_{\rm sub}W_{\rm sub}) =6.51$~m \\
        Radial velocity resolution & $\lambda_{o}/(2 N_{\rm sym}T_{\rm sym}) =9.37$~m/s  \\
        \rowcolor{Gray} Azimuth resolution \newline($3$-dB beamwidth) & \mbox{}\newline $10^\circ$ \\
        Elevation resolution \newline($3$-dB beamwidth) & \mbox{}\newline $10^\circ$\\
        \rowcolor{Gray} Maximum range allowed by the \newline cyclic prefix & \mbox{}\newline $c_{o}T_{o}/2 = 713.79$~m\\
        Maximum non-ambiguous range & $c_{o}/(2W_{\rm sub})=208.19$~m \\
        \rowcolor{Gray} Maximum non-ambiguous radial\newline velocity (at the carrier frequency) & \mbox{}\newline $\lambda_{o}/(4T_{\rm sym})=299.79$~m/s\\
        Inspected range interval & $[R_{\min},R_{\max}] = [10~\text{m},\,200~\text{m}]$ \\
        \rowcolor{Gray} Inspected azimuth interval& $[\theta_{\min}^{\rm{az}},\theta_{\max}^{\rm{az}}]=[-22.5^\circ,\, 22.5^\circ]$ \\
        Inspected elevation interval & $[\theta_{\min}^{\rm{el}},\theta_{\max}^{\rm{el}}]=[5^\circ,\, 15^\circ]$\\
        \rowcolor{Gray} Number of delay sampling points & $N_{\rm del}=60$ \\
        Number of Doppler sampling \newline points & \mbox{}\newline $N_{\rm dop} =9$\\
        \rowcolor{Gray} Number of azimuth pointing \newline directions & \mbox{}\newline $N_{\rm dir}^{\rm az}=6$\\
        Number of elevation pointing \newline directions & \mbox{}\newline $N_{\rm dir}^{\rm el}=1$\\ 
        \rowcolor{Gray} Scan duration & $T_{\rm scan}=0.06$~s\\
        \bottomrule    
    \end{tabular}
    \end{center}
\end{table}

Finally, in the correlators of the radar front-end, the  weights $\{w_{q}(n)\}$ are set all equal; in the detector and plot-extractor, the threshold $\eta_{\rm plot}$ is set to $5$, which results in an average number of plots per scan equal to $6$ under hypothesis $\mathcal H_0$; in the TBD processor, a polynomial regression is used to smooth the trajectory of detected targets and improve the measurement accuracy, a maximum target speed of $40$~m/s is assumed, and the threshold $\eta_{\rm TBD}$ is set to have $P_{\rm fa} =10^{-3}$. 

\subsection{Numerical results}\label{subsec_num_res}

The communication performance is evaluated in terms of user sum rate $\mathcal R$ in~\eqref{user-sum-rate}, while the radar performance in terms of the probability of detection ($P_{\rm d}$, which is the probability of declaring $\mathcal{H}_1$ under $\mathcal{H}_1$) and the root mean square error (RMSE) on the position estimation. For comparison, different number of users ($K=3,6$) and scans ($N_{\rm scan}=1,5,8,12,15)$ are considered.

\begin{figure}[!t]
    \centering
    \includegraphics[width=0.9\columnwidth]{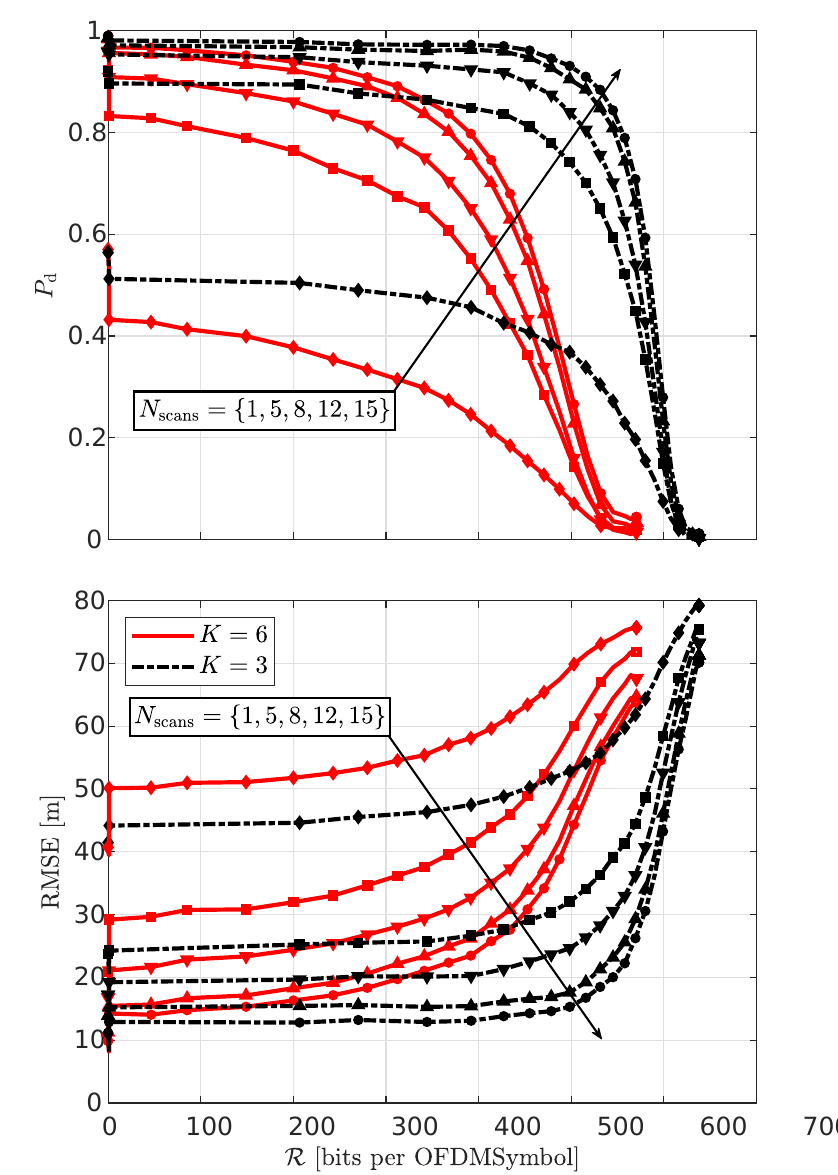}
    \caption{Probability of target detection (top) and RMSE on the position estimation (bottom) versus the user sum-rate for various numbers of users and scans. Here, the number of scans is $N_{\rm{scans}}=1,5,8,12,15$, which increases in the direction of the arrow and corresponds to markers \scalebox{0.85}{$\blacklozenge$}, \scalebox{0.75}{$\blacksquare$}, $\blacktriangledown$, $\blacktriangle$, \scalebox{1.25}{$\bullet$}, respectively.}
    \label{fig:sr_pd}
\end{figure}

Fig.~\ref{fig:sr_pd} shows $P_{\rm d}$ and RMSE versus $\mathcal R$. Each curve is obtained by varying the fraction of power $\gamma_r$ devoted to the radar function and reporting the corresponding pair $(\mathcal R, P_{\rm d})$; therefore, it 
can be interpreted as a \emph{system operating characteristic}. It is seen by inspection that $P_{\rm d}$ improves when the number of scans processed by the radar detector increases and when the number of users decreases. In particular, notice here the significant gain granted by the TBD processor: e.g., when $K=3$, the achievable $\mathcal R$ for $P_{\rm d}=0.5$ is 225~bits/symbol when $N_{\rm scan}=1$, that grows to $572-590$~bits/symbol when $N_{\rm scan}$ is increased to $5-15$. It is also interesting to notice that $P_{\rm d}$ is independent of $K$ when $\mathcal R=0$, since $\gamma_r=1$ and the communication function is not active, and that $P_{\rm d}$ for $\mathcal R=0$ is greater than $P_{\rm d}$ for $\mathcal R\rightarrow 0^+$ since, in this latter case, the communication function is active, whereby the radar beamformers must satisfy the orthogonality constraint ($\bm H_q\herm \bm f_{q,r}=\bm 0_K$ for $q=1, \ldots, N_{\rm sub}$), which reduces the degrees of freedom for system optimization. Furthermore, when all the power is devoted to the communication function ($\gamma_r=0$), $P_{\rm d}$ is still larger than $P_{\rm fa}$ since part of the communication signal reaches the RIS and, therefore, can be used for target detection. Similar considerations hold for the RMSE; in particular, notice the large gain in terms of estimation accuracy granted by the multi-frame processing: e.g., when the number of processed scans is increased from 1 to 15, the RMSE for $\mathcal R=300$~bits per OFDM symbol decreases from 45.9~m (54.0~m) to 13.0~m (19.2~m) with 3 users (6 users).

\begin{figure}[!t]
    \centering
    \includegraphics[width=0.9\columnwidth]{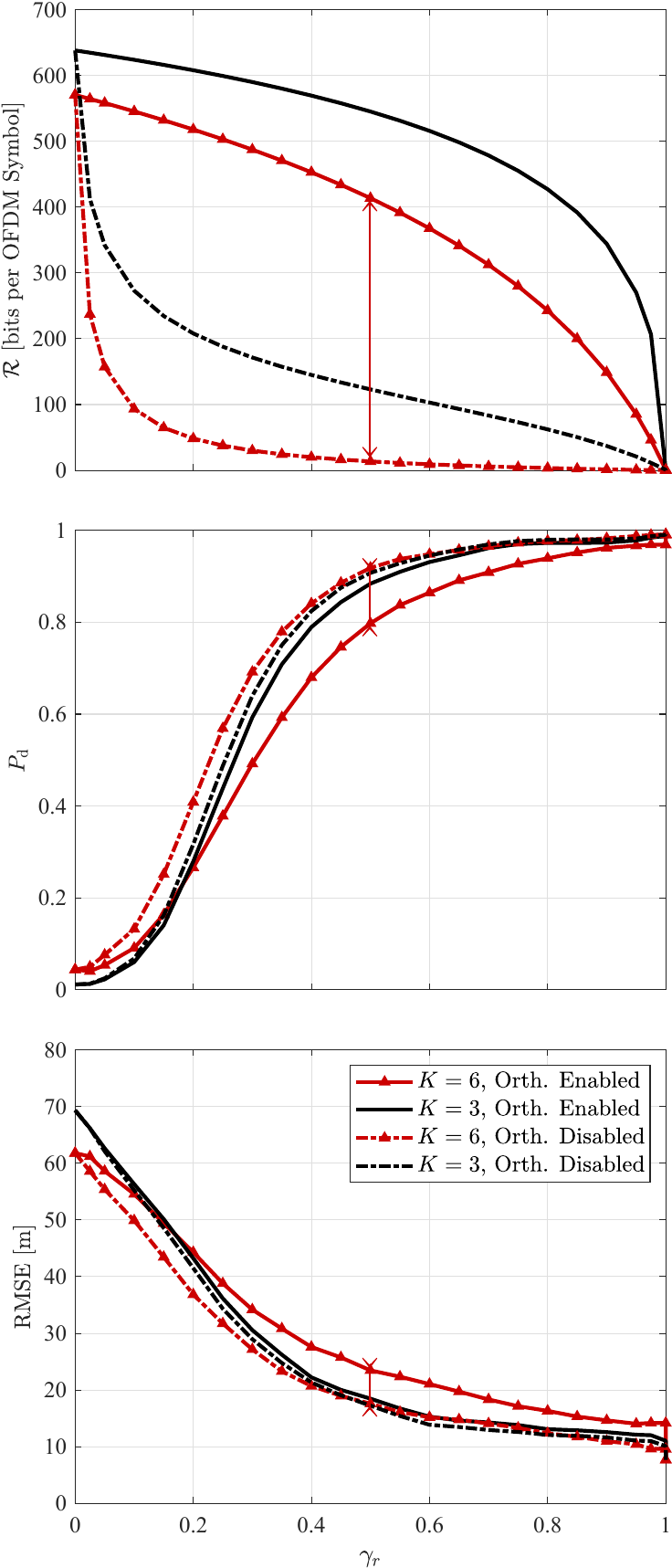}
    \caption{User sum-rate (top), probability of target detection (middle), and RMSE in the position estimation (bottom) versus the fraction of power devoted to the radar function for different numbers of users when $N_{\rm{scans}} = 15$~scans are processed, and the orthogonality constraint in~\eqref{orth_constraint} enforced on the radar beamformers is enabled/disabled.}
    \label{fig:gamma_all}
\end{figure}

Fig.~\ref{fig:gamma_all} verifies the impact of the orthogonality constraint in~\eqref{orth_constraint} enforced on the radar beamformers. In particular, $\mathcal R$, $P_{\rm d}$, and RMSE are plotted versus $\gamma_r$ for different numbers of users when $N_{\rm{scans}}=15$; the solid/dashed lines correspond to the case where the orthogonality constraint in~\eqref{orth_constraint} is enabled/disabled, respectively. As it can be seen, the removal of this constraint results in a small gain in terms of $P_{\rm d}$ and RMSE and, at the same time, in a significant loss in terms of achievable user sum rate caused by the radar interference; e.g., for $\gamma_{r} = 0.5$ and $K=6$, the radar performance only marginally improves by disabling the orthogonal constraint, with $P_{\rm d}$ increasing from 0.8 to 0.91 and RMSE reducing from 23.5~m to 17.6~m, while the user sum rate significantly drops from 413~bits/symbol to 13~bits/symbol.

\begin{figure}[t!]
    \centering
    \includegraphics[width=0.9\columnwidth]{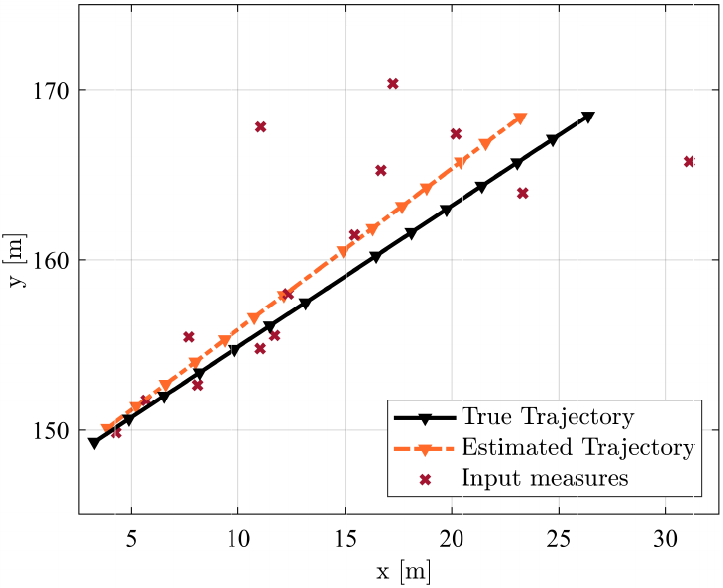}
    \caption{One instance of the true and estimated trajectory of the target, along with the plots passed to the TBD processor when $N_{\rm{scans}} = 15$.}
    \label{fig:trajectory}
\end{figure}

Finally, Fig.~\ref{fig:trajectory} reports an instance of the true and estimated trajectory of the target, along with the plots passed to the TBD processor when $N_{\rm{scans}} = 15$. It is seen by inspection the large improvement in the estimation accuracy granted by the multi-frame processing, which is able to average out and significantly reduce the large uncertainty in the input measures due to the poor angular resolution.

\section{Conclusions}\label{sec:conclusions}

In this paper, we have considered a BS accomplishing both downlink communications towards multiple ground users and RIS-aided monostatic radar sensing towards the sky. Upon relying on the OFDM modulation format employed in current 5G systems, we have developed a convenient signal model and introduced the definition of the transmit, receive, and two-way beampatterns of the RIS-assisted monostatic radar. We have then considered a modular approach for system design, in which transmit beamformers dedicated to the sensing and communication functions are independently designed. In this context, we have proposed a novel method to design the transmit beamformer dedicated to the radar, aimed at controlling the sidelobe levels of the two-way beampattern of the RIS-assisted radar in the presence of eavesdroppers, jammers, and other scattering objects, while avoiding any radar interference to the users. Also, we have proposed employing a multi-frame detector at the radar receiver to exploit the time correlation among the echoes generated by a moving target over consecutive scans; the considered radar detector includes a TBD processor in addition to the canonical detector and plot-extractor. A major result of this study is that, by increasing the number of scans jointly processed by the radar detector (and therefore by increasing its implementation complexity), we can reduce the power dedicated to the radar function while maintaining the same sensing performance (measured in terms of probability of detection and root mean square error in the position estimation); this excess power can then be reused to increase the user sum-rate. Hence, more favorable sensing and communication tradeoffs are possible.

Future works should account for the presence of estimation errors in the user channels. Also, they could further investigate the potential benefits of resorting to a multi-frame processing when the sensing and communication transmit beamformers are jointly designed, when the BS is mounted on a moving platform, or when multiple BSs cover the same region. In particular, the multi-frame radar processing might also improve the localization accuracy of legacy mobile users, thus helping the construction of more efficient schemes for resource scheduling and for channel estimation and prediction.  Finally, 
an experimental validation assessing the achievable performance under real-world operating conditions is still missing and should be verified and discussed in future studies.

\bibliographystyle{IEEEtran}
\bibliography{references}

\begin{thebibliography}{10}
\providecommand{\url}[1]{#1}
\csname url@samestyle\endcsname
\providecommand{\newblock}{\relax}
\providecommand{\bibinfo}[2]{#2}
\providecommand{\BIBentrySTDinterwordspacing}{\spaceskip=0pt\relax}
\providecommand{\BIBentryALTinterwordstretchfactor}{4}
\providecommand{\BIBentryALTinterwordspacing}{\spaceskip=\fontdimen2\font plus
\BIBentryALTinterwordstretchfactor\fontdimen3\font minus
  \fontdimen4\font\relax}
\providecommand{\BIBforeignlanguage}[2]{{%
\expandafter\ifx\csname l@#1\endcsname\relax
\typeout{** WARNING: IEEEtran.bst: No hyphenation pattern has been}%
\typeout{** loaded for the language `#1'. Using the pattern for}%
\typeout{** the default language instead.}%
\else
\language=\csname l@#1\endcsname
\fi
#2}}
\providecommand{\BIBdecl}{\relax}
\BIBdecl

\bibitem{9606831}
Y.~Cui, F.~Liu, X.~Jing, and J.~Mu, ``{Integrating Sensing and Communications
  for Ubiquitous IoT: Applications, Trends, and Challenges},'' \emph{IEEE
  Network}, vol.~35, no.~5, pp. 158--167, 2021.

\bibitem{9737357}
F.~Liu, Y.~Cui, C.~Masouros, J.~Xu, T.~X. Han, Y.~C. Eldar, and S.~Buzzi,
  ``{Integrated Sensing and Communications: Toward Dual-Functional Wireless
  Networks for 6G and Beyond},'' \emph{IEEE Journal on Selected Areas in
  Communications}, vol.~40, no.~6, pp. 1728--1767, 2022.

\bibitem{9705498}
A.~Liu, Z.~Huang, M.~Li, Y.~Wan, W.~Li, T.~X. Han, C.~Liu, R.~Du, D.~K.~P. Tan,
  J.~Lu, Y.~Shen, F.~Colone, and K.~Chetty, ``{A Survey on Fundamental Limits
  of Integrated Sensing and Communication},'' \emph{IEEE Communications Surveys
  \& Tutorials}, vol.~24, no.~2, pp. 994--1034, 2022.

\bibitem{10012421}
Z.~Wei, H.~Qu, Y.~Wang, X.~Yuan, H.~Wu, Y.~Du, K.~Han, N.~Zhang, and Z.~Feng,
  ``{Integrated Sensing and Communication Signals Toward 5G-A and 6G: A
  Survey},'' \emph{IEEE Internet of Things Journal}, vol.~10, no.~13, pp.
  11\,068--11\,092, 2023.

\bibitem{9945983}
F.~Dong, F.~Liu, Y.~Cui, W.~Wang, K.~Han, and Z.~Wang, ``{Sensing as a Service
  in 6G Perceptive Networks: A Unified Framework for ISAC Resource
  Allocation},'' \emph{IEEE Transactions on Wireless Communications}, vol.~22,
  no.~5, pp. 3522--3536, 2023.

\bibitem{10418473}
S.~Lu, F.~Liu, Y.~Li, K.~Zhang, H.~Huang, J.~Zou, X.~Li, Y.~Dong, F.~Dong,
  J.~Zhu, Y.~Xiong, W.~Yuan, Y.~Cui, and L.~Hanzo, ``{Integrated Sensing and
  Communications: Recent Advances and Ten Open Challenges},'' \emph{IEEE
  Internet of Things Journal}, vol.~11, no.~11, pp. 19\,094--19\,120, 2024.

\bibitem{9965407}
M.~Liu, M.~Yang, H.~Li, K.~Zeng, Z.~Zhang, A.~Nallanathan, G.~Wang, and
  L.~Hanzo, ``{Performance Analysis and Power Allocation for Cooperative ISAC
  Networks},'' \emph{IEEE Internet of Things Journal}, vol.~10, no.~7, pp.
  6336--6351, 2023.

\bibitem{10623531}
Z.~Ni, J.~A. Zhang, X.~Huang, and R.~P. Liu, ``{Frequency-Time Resource
  Allocation for Multiuser Uplink ISAC Systems},'' \emph{IEEE Transactions on
  Vehicular Technology}, vol.~73, no.~12, pp. 18\,893--18\,906, 2024.

\bibitem{10694524}
G.~Mylonopoulos, B.~Makki, G.~Fodor, and S.~Buzzi, ``{Extended ARQ Protocol For
  Reliable Integrated Sensing and Communication Systems},'' in \emph{2024 IEEE
  25th International Workshop on Signal Processing Advances in Wireless
  Communications (SPAWC)}, 2024, pp. 321--325.

\bibitem{9557830}
W.~Yuan, Z.~Wei, S.~Li, J.~Yuan, and D.~W.~K. Ng, ``{Integrated Sensing and
  Communication-Assisted Orthogonal Time Frequency Space Transmission for
  Vehicular Networks},'' \emph{IEEE Journal of Selected Topics in Signal
  Processing}, vol.~15, no.~6, pp. 1515--1528, 2021.

\bibitem{9540344}
J.~A. Zhang, F.~Liu, C.~Masouros, R.~W. Heath, Z.~Feng, L.~Zheng, and
  A.~Petropulu, ``{An Overview of Signal Processing Techniques for Joint
  Communication and Radar Sensing},'' \emph{IEEE Journal of Selected Topics in
  Signal Processing}, vol.~15, no.~6, pp. 1295--1315, 2021.

\bibitem{9765815}
M.~Jian, G.~C. Alexandropoulos, E.~Basar, C.~Huang, R.~Liu, Y.~Liu, and
  C.~Yuen, ``{Reconfigurable intelligent surfaces for wireless communications:
  Overview of hardware designs, channel models, and estimation techniques},''
  \emph{Intelligent and Converged Networks}, vol.~3, no.~1, pp. 1--32, 2022.

\bibitem{10736517}
G.~Mylonopoulos, B.~Makki, S.~Buzzi, and G.~Fodor, ``{Joint User Detection and
  Localization in Near-Field Using Reconfigurable Intelligent Surfaces},''
  \emph{IEEE Wireless Communications Letters}, vol.~14, no.~1, pp. 58--62,
  2025.

\bibitem{10077119}
R.~Liu, M.~Li, H.~Luo, Q.~Liu, and A.~L. Swindlehurst, ``{Integrated Sensing
  and Communication with Reconfigurable Intelligent Surfaces: Opportunities,
  Applications, and Future Directions},'' \emph{IEEE Wireless Communications},
  vol.~30, no.~1, pp. 50--57, 2023.

\bibitem{9852716}
H.~Luo, R.~Liu, M.~Li, Y.~Liu, and Q.~Liu, ``{Joint Beamforming Design for
  RIS-Assisted Integrated Sensing and Communication Systems},'' \emph{IEEE
  Transactions on Vehicular Technology}, vol.~71, no.~12, pp. 13\,393--13\,397,
  2022.

\bibitem{9844707}
H.~Zhang, ``{Joint Waveform and Phase Shift Design for RIS-Assisted Integrated
  Sensing and Communication Based on Mutual Information},'' \emph{IEEE
  Communications Letters}, vol.~26, no.~10, pp. 2317--2321, 2022.

\bibitem{10052711}
H.~Luo, R.~Liu, M.~Li, and Q.~Liu, ``{RIS-Aided Integrated Sensing and
  Communication: Joint Beamforming and Reflection Design},'' \emph{IEEE
  Transactions on Vehicular Technology}, vol.~72, no.~7, pp. 9626--9630, 2023.

\bibitem{10792983}
P.~Saikia, A.~Jee, K.~Singh, C.~Pan, W.-J. Huang, and T.~A. Tsiftsis,
  ``Ris-aided integrated sensing and communication systems: Star-ris versus
  passive ris?'' \emph{IEEE Open Journal of the Communications Society},
  vol.~5, pp. 7954--7973, 2024.

\bibitem{10466748}
Q.~Li, M.~El-Hajjar, and L.~Hanzo, ``Ergodic spectral efficiency analysis of
  intelligent omni-surface aided systems suffering from imperfect csi and
  hardware impairments,'' \emph{IEEE Transactions on Communications}, vol.~72,
  no.~8, pp. 5073--5086, 2024.

\bibitem{10156858}
Q.~Li, M.~El-Hajjar, Y.~Sun, I.~Hemadeh, A.~Shojaeifard, Y.~Liu, and L.~Hanzo,
  ``Achievable rate analysis of the star-ris-aided noma uplink in the face of
  imperfect csi and hardware impairments,'' \emph{IEEE Transactions on
  Communications}, vol.~71, no.~10, pp. 6100--6114, 2023.

\bibitem{10025392}
Q.~Li, M.~El-Hajjar, I.~Hemadeh, D.~Jagyasi, A.~Shojaeifard, and L.~Hanzo,
  ``Performance analysis of active ris-aided systems in the face of imperfect
  csi and phase shift noise,'' \emph{IEEE Transactions on Vehicular
  Technology}, vol.~72, no.~6, pp. 8140--8145, 2023.

\bibitem{10915665}
P.~Saikia, A.~Jee, K.~Singh, W.-J. Huang, A.-A.~A. Boulogeorgos, and T.~A.
  Tsiftsis, ``Hybrid-ris empowered uav-assisted isac systems: Transfer
  learning-based drl,'' \emph{IEEE Transactions on Communications}, pp. 1--1,
  2025.

\bibitem{Boers_2004}
Y.~Boers and J.~N. Driessen, ``{Multitarget Particle Filter Track Before Detect
  Application},'' \emph{{IEE} Proc. Radar Sonar Navig.}, vol. 151, no.~6, pp.
  351--357, Dec. 2004.

\bibitem{Pulford_2010}
G.~W. Pulford and B.~F.~L. Scala, ``{Multihypothesis Viterbi Data Association:
  Algorithm Development and Assessment},'' \emph{{IEEE} Trans. Aerosp.
  Electron. Syst.}, vol.~46, no.~2, pp. 583--609, Apr. 2010.

\bibitem{Blanding_2007}
W.~R. Blanding, P.~K. Willett, Y.~Bar-Shalom, and R.~S. Lynch, ``{Multiple
  Target Tracking Using Maximum Likelihood Probabilistic Data Association},''
  in \emph{{IEEE} Aerosp. Conf.}, Big {S}ky, {MT}, {USA}, Mar. 2007.

\bibitem{Davey_2008}
S.~J. Davey, M.~G. Rutten, and B.~Cheung, ``{A comparison of Detection
  Performance for Several Track-Before-Detect Algorithms},'' in
  \emph{International Conference on Information Fusion ({FUSION})}, Cologne,
  Germany, 2008.

\bibitem{Grossi-2013a}
E.~Grossi, M.~Lops, and L.~Venturino, ``{A Novel Dynamic Programming Algorithm
  for Track-Before-Detect in Radar Systems},'' \emph{IEEE Transactions on
  Signal Processing}, vol.~61, no.~10, pp. 2608--2619, May 2013.

\bibitem{Grossi-2013b}
------, ``{A Track-Before-Detect Algorithm With Thresholded Observations and
  Closely-Spaced Targets},'' \emph{IEEE Signal Processing Letters}, vol.~20,
  no.~12, pp. 1171--1174, Dec. 2013.

\bibitem{Aprile-2016}
A.~Aprile, E.~Grossi, M.~Lops, and L.~Venturino, ``{Track-Before-Detect for Sea
  Clutter Rejection: Tests with Real Data},'' \emph{IEEE Transactions on
  Aerospace and Electronic Systems}, vol.~52, no.~3, pp. 1035--1045, Jun. 2016.

\bibitem{ji2024modified}
W.~Ji, T.~Liu, Y.~Song, H.~Yin, B.~Tian, and N.~Zhu, ``{Modified Hybrid
  Integration Algorithm for Moving Weak Target in Dual-Function Radar and
  Communication System},'' \emph{Remote Sensing}, vol.~16, no.~19, p. 3601,
  2024.

\bibitem{9591331}
X.~Wang, Z.~Fei, J.~Huang, and H.~Yu, ``{Joint Waveform and Discrete Phase
  Shift Design for RIS-Assisted Integrated Sensing and Communication System
  Under Cramer-Rao Bound Constraint},'' \emph{IEEE Transactions on Vehicular
  Technology}, vol.~71, no.~1, pp. 1004--1009, 2022.

\bibitem{9729741}
Y.~He, Y.~Cai, H.~Mao, and G.~Yu, ``{RIS-Assisted Communication Radar
  Coexistence: Joint Beamforming Design and Analysis},'' \emph{IEEE Journal on
  Selected Areas in Communications}, vol.~40, no.~7, pp. 2131--2145, 2022.

\bibitem{9858656}
K.~Meng, Q.~Wu, S.~Ma, W.~Chen, K.~Wang, and J.~Li, ``{Throughput Maximization
  for UAV-Enabled Integrated Periodic Sensing and Communication},'' \emph{IEEE
  Transactions on Wireless Communications}, vol.~22, no.~1, pp. 671--687, 2023.

\bibitem{10158711}
Z.~He, W.~Xu, H.~Shen, D.~W.~K. Ng, Y.~C. Eldar, and X.~You, ``{Full-Duplex
  Communication for ISAC: Joint Beamforming and Power Optimization},''
  \emph{IEEE Journal on Selected Areas in Communications}, vol.~41, no.~9, pp.
  2920--2936, 2023.

\bibitem{9668964}
Z.~Wang, Y.~Liu, X.~Mu, Z.~Ding, and O.~A. Dobre, ``{NOMA Empowered Integrated
  Sensing and Communication},'' \emph{IEEE Communications Letters}, vol.~26,
  no.~3, pp. 677--681, 2022.

\bibitem{10086626}
H.~Hua, J.~Xu, and T.~X. Han, ``{Optimal Transmit Beamforming for Integrated
  Sensing and Communication},'' \emph{IEEE Transactions on Vehicular
  Technology}, vol.~72, no.~8, pp. 10\,588--10\,603, 2023.

\bibitem{Buzzi-2022}
S.~Buzzi, E.~Grossi, M.~Lops, and L.~Venturino, ``{Foundations of MIMO Radar
  Detection Aided by Reconfigurable Intelligent Surfaces},'' \emph{IEEE
  Transactions on Signal Processing}, vol.~70, pp. 1749--1763, 2022.

\bibitem{10360201}
S.~Kumar, A.~Jee, and S.~Prakriya, ``Performance analysis and optimization of
  partitioned-irs-assisted noma network,'' \emph{IEEE Wireless Communications
  Letters}, vol.~13, no.~3, pp. 761--765, 2024.

\bibitem{Van-Trees-IV}
H.~L. {Van Trees}, \emph{{Detection, Estimation, and Modulation Theory, {Part
  IV}: Optimum Array Processing}}.\hskip 1em plus 0.5em minus 0.4em\relax New
  York, NY, USA: John Wiley \& Sons, 2002.

\bibitem{book-Stutzman03}
W.~L. Stutzman and G.~A. Thiele, \emph{{Antenna Theory and Design}},
  3rd~ed.\hskip 1em plus 0.5em minus 0.4em\relax New York, NY, USA: John Wiley
  \& Sons, 1998.

\bibitem{zf_methods}
Q.~Spencer, A.~Swindlehurst, and M.~Haardt, ``{Zero-Forcing Methods for
  Downlink Spatial Multiplexing in Multiuser MIMO Channels},'' \emph{IEEE
  Transactions on Signal Processing}, vol.~52, no.~2, pp. 461--471, 2004.

\bibitem{Venturino-2015}
L.~Venturino and S.~Buzzi, ``{Energy-Aware and Rate-Aware Heuristic Beamforming
  in Downlink MIMO OFDMA Networks With Base-Station Coordination},'' \emph{IEEE
  Transactions on Vehicular Technology}, vol.~64, no.~7, pp. 2897--2910, Jul.
  2015.

\bibitem{10309244}
A.~Jee and S.~Prakriya, ``Novel channel aware power control for a multi-user
  downlink noma network,'' \emph{IEEE Wireless Communications Letters},
  vol.~13, no.~2, pp. 392--396, 2024.

\bibitem{Skolnik-book}
M.~I. Skolnik, \emph{{Introduction to RADAR Systems}}, 3rd~ed.\hskip 1em plus
  0.5em minus 0.4em\relax New York, NY, USA: McGraw-Hill Higher Education,
  2015.

\bibitem{Chen-2023}
X.~Chen, J.~An, Z.~Xiong, C.~Xing, N.~Zhao, F.~R. Yu, and A.~Nallanathan,
  ``{Covert Communications: A Comprehensive Survey},'' \emph{IEEE
  Communications Surveys \& Tutorials}, vol.~25, no.~2, pp. 1173--1198,
  Secondquarter 2023.

\bibitem{Richards-book}
M.~A. Richards, \emph{{Foundamentals of RADAR Signal Processing}}.\hskip 1em
  plus 0.5em minus 0.4em\relax New York, NY, USA: McGraw-Hill Higher Education,
  2005.

\bibitem{Bertsekas_1999}
D.~P. Bertsekas, \emph{{Nonlinear Programming}}, 2nd~ed.\hskip 1em plus 0.5em
  minus 0.4em\relax Belmont, {MA}, {USA}: Athena Scientific, 1999.

\bibitem{Nocedal_2006}
J.~Nocedal and S.~Wright, \emph{{Numerical Optimization}}, 2nd~ed.\hskip 1em
  plus 0.5em minus 0.4em\relax New York, {NY}, {USA}: Springer, 2006.

\bibitem{9223720}
P.~Xu, G.~Chen, Z.~Yang, and M.~D. Renzo, ``{Reconfigurable Intelligent
  Surfaces-Assisted Communications With Discrete Phase Shifts: How Many
  Quantization Levels Are Required to Achieve Full Diversity?}'' \emph{IEEE
  Wireless Communications Letters}, vol.~10, no.~2, pp. 358--362, 2021.

\end{thebibliography}

\end{document}